\documentclass[11pt]{article}

\usepackage[margin=1in]{geometry}
\usepackage{amsmath,amssymb,amsfonts}
\usepackage{graphicx}
\usepackage{booktabs}
\usepackage{natbib}
\usepackage{hyperref}
\usepackage{xcolor}
\usepackage{algorithm}
\usepackage{algorithmic}
\usepackage{enumitem}
\usepackage{bm}
\usepackage{float}

\newcommand{\vect}[1]{\bm{#1}}

\DeclareMathOperator{\softplus}{softplus}
\DeclareMathOperator{\sigmoid}{sigmoid}
\DeclareMathOperator{\logit}{logit}

\title{High-resolution Calibrated Probabilistic Hourly Precipitation from a Deterministic Forecast}

\author{Thomas M.\ Hamill\thanks{The Weather Company, Atlanta,
  Georgia; e-mail: \texttt{tom.hamill@weather.com}}}

\date{\today}

\begin{document}

\maketitle

\begin{abstract}
An  ``Attention Residual U-Net'' method is described for probabilistic quantitative
precipitation forecasting (PQPF) that predicts the hourly probability of no precipitation plus
the distribution of positive precipitation from a weighted mixture of two Gamma distributions.
The neural network is trained on patches of numerical weather prediction
(NWP) hourly precipitation from The Weather Company's convection-permitting GRAF (Global
high-Resolution Atmospheric Forecasting) model along with terrain information and
column-average relative humidity from the National Oceanic and
Atmospheric Administration's (NOAA's) Global Forecast System (GFS).
The target data are NOAA's Multi-Radar, Multi-Sensor (MRMS)
gauge-corrected, quality controlled radar data sampled to the same grid as the GRAF data.
The network outputs distributional parameters for each model grid point.
Training uses negative log-likelihood as a proper scoring rule,
with climatological initialization for stable convergence.
Inference is performed as a single forward pass over the contiguous United States (CONUS)
domain, with edge-replication padding to satisfy the network's spatial-divisibility
requirement.  The subsequent
forecasts are spatially detailed, highly reliable, and skillful with respect to climatology and a simpler
reference forecast method.  The method is particularly useful for estimating probabilities 
in regions with large terrain variation.
\end{abstract}

\section{Introduction}
\label{sec:intro}

High-resolution probabilistic precipitation forecasts can be helpful
for flash-flood warnings, hydrological modeling, and decision
support.  The choice and effective use of probabilistic
guidance for operational forecasters and emergency managers is
itself an active research area, e.g.,
\citep{murphy1993, gneiting2007, hoss2016, raftery2016, cross2021,
ripberger2022, heggli2023, radford2023}.

Generating reliable probabilistic forecast content commonly is complicated by
model errors and computational resource limitations.
Raw numerical weather prediction (NWP) precipitation
fields exhibit systematic biases in both location and intensity, and
a single deterministic forecast cannot convey forecast uncertainty.
This motivates the use of ensemble forecasts, in some situations
with accompanying statistical postprocessing.
However, for convection-permitting forecasts over large domains with
conventional NWP models, ensemble forecasts may be too computationally expensive to
perform regularly.    Is there a way to provide high-quality, unbiased probabilistic
precipitation guidance from a single deterministic forecast?

A simple method to estimate probabilities may be to incorporate relevant data from the
surrounding neighborhood.   This can be applied,
for example, by estimating the probability of exceedance using relative frequency
derived from gridded forecast data within some distance \citep{theis2005}.  This can further be improved
by applying distance-dependent weights.

Traditional statistical post-processing methods may also be used to generate probabilities
from deterministic forecasts by fitting parametric distributions based on
forecast/observation pairs.  Model Output Statistics
\citep[MOS;][]{glahn1972, carter1989} pioneered regression-based
calibration of NWP output to produce probabilistic guidance.
There is now a burgeoning literature on the use and calibration of
ensembles for generation of probabilities, including extended
logistic regression \citep{bentzien2012, benbouall2013},
heteroscedastic regression \citep{messner2014}, probabilistic ensemble
model output statistics \citep[EMOS;][]{scheuerer2014, scheuerer2015}, ensemble
model output statistics \citep{hamill2004}, and Bayesian model
averaging specifically designed for probabilistic quantitative
precipitation forecasting \citep[BMA--PQPF;][]{sloughter2007}, and many more.

National-scale blending of statistically post-processed forecasts
\citep{hamill2017} and neural-network-based post-processing
\citep{hamill2018, ghazvinian2022} have further extended the
capability.  These methods have demonstrated clear improvements in
probabilistic skill, but they typically operate on a grid-point or
station basis, exploiting neither the spatial structure of the NWP
fields nor the relationships among neighboring forecast errors.

Deep learning offers a natural extension that can exploit
spatial context.  Convolutional neural network based 
approaches \citep[CNNs;][]{lecun1998} are
particularly well suited to precipitation post-processing because
their convolutional filters efficiently capture local spatial
patterns in forecast errors, and
their hierarchical structure can represent corrections
simultaneously at multiple spatial scales.  In the context of
probabilistic prediction, they may extract scale-dependent
predictive information.  Examples of recent deep learning
applications for probabilistic precipitation post-processing
include \citet{chapman2022}, \citet{badrinath2023}, and
\citet{hu2023}.

The work described here leverages a type of convolutional 
neural network, but it is designed to produce a probabilistic
forecast from a deterministic one, in this case The Weather Company's GRAF model (Section~\ref{sec:graf}).
This is a conventional convection-permitting NWP system based on the 
National Center for Atmospheric Research (NCAR) Model for Predictions Across 
Scales \citep[MPAS;][]{skamarock2012}.  The method is an \emph{Attention Residual U-Net}
--- a U-shaped encoder--decoder architecture augmented with
residual connections and attention gates. It predicts the
parameters of a \emph{zero-inflated mixture of two Gamma
distributions} at each grid point, i.e., a point mass probability
of zero precipitation and a weighted mixture of two Gamma distributions.  The two-component Gamma
mixture provides greater flexibility than a single Gamma,
accommodating for example both the light-precipitation mode associated with
stratiform rain and the heavy-precipitation tail associated with
convective events.  By incorporating terrain information as predictive features, the probabilistic forecasts
that it generates can be linked to terrain variations, which we will show has predictive advantages
over a simpler method such as a neighborhood method \citep{theis2005, schwartz2017}.

A convection-permitting model like GRAF is capable of producing forecast precipitation at high
temporal and spatial resolution.  Hence, a probabilistic postprocessing method ideally ought
to be able to produce guidance with similar spatial and temporal resolution.
Accordingly, this study will focus on {\it hourly} precipitation
accumulation on a 4-km grid, a more challenging problem than accumulation
over longer time spans and larger regions \citep{roberts2008}.

The rest of the manuscript is organized as follows.   Section 2 briefly describes the GRAF model
underpinning this study.  Section 3 describes the deep neural network, including the data used, the
training and inference methodology, and the loss function.  Section 4 briefly describes the methodology
used to produce the neighborhood reference forecasts.   Section 5 describes the verification metrics
used to assess forecast quality.  Section 6 provides forecast case studies and verification results,
and section 7 provides conclusions and possible next steps.

\section{The GRAF Forecast Model}
\label{sec:graf}

\begin{figure}[H]
\centering
\includegraphics[width=0.85\textwidth]{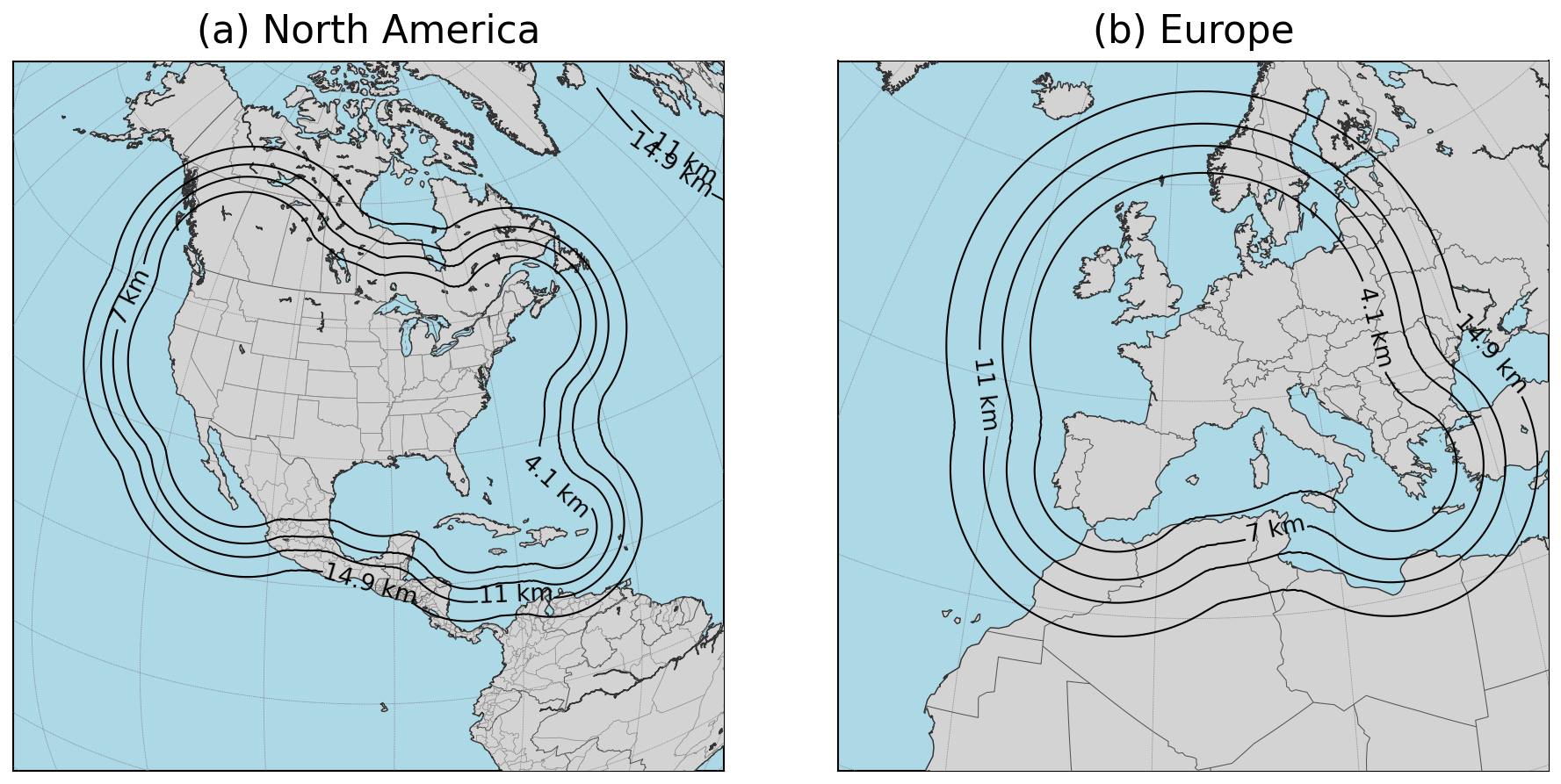}
\caption{Variable mesh spacing for the GRAF model over (a) North America,
and (b) Europe.  Outside these areas, mesh spacing is 14-15 km.}
\label{fig:graf_mesh}
\end{figure}

The Weather Company (TWCo) GRAF (Global high-Resolution Atmospheric
Forecasting) model is based on version 6.3 of the NCAR Model for
Predictions Across Scales \citep[MPAS;][]{skamarock2012}, designed to run
efficiently on graphical processing units (GPUs).
GRAF uses a variable-resolution unstructured centroidal-Voronoi mesh
that achieves approximately 4\,km grid spacing over the contiguous
United States (CONUS), Caribbean Basin, and western Europe,
relaxing to approximately 15\,km elsewhere (Fig.~\ref{fig:graf_mesh}).
The design provides the convenience of a global forecast where lateral boundary 
conditions are not needed, yet still providing forecasts capable of modeling
organized thunderstorms.   In 2026, GRAF makes forecasts every hour, with even-hour
Coordinated Universal Time (UTC) initial-time forecasts to +72 h and odd-hour forecasts to +15 h.   Initial conditions
are generated for each hour, using the most recently available European Centre for Medium-Range Weather Forecasts (ECMWF) analysis and then
running a 3-D variational analysis procedure \citep{lorenc1986} using the Joint Effort for Data assimilation
Integration (JEDI) software \citep{tremolet2020, ha2022, park2023} to update to recently available surface observations, rawinsonde,
and radar reflectivity data.  GRAF data are disseminated to media, aviation,
consumer, agriculture, and energy customers.
For this study, we will consider the calibration of probabilities only to +48 h lead time.

GRAF employs a scale-aware nTiedtke
convective parameterization \citep{wang2022} that transitions smoothly
from full convective parameterization at 15\,km toward near-explicit
convection at 4\,km.  Additional parameterizations include the
Yonsei University (YSU) planetary boundary layer scheme \citep{hong2006ysu} with gravity-wave drag,
Weather Research and Forecasting (WRF) single-moment (WSM6)
microphysics \citep{hong2006} with Thompson cloud fraction, the
Rapid Radiative Transfer Model for General Circulation Models
\citep[RRTMG;][]{mlawer1997,clough2005}, and the
NOAH 4-layer land surface model \citep{ek2003}.  Terrain heights are
from the Global Multi-resolution Terrain Elevation Data 2010
(GMTED2010), with albedo and land-characteristic data from
MODIS (Moderate Resolution Imaging Spectroradiometer) at
30\,arcsec grid spacing.

Before use in the Attention ResUNet, GRAF forecasts from the native mesh
are interpolated to a 4-km grid covering the CONUS, southern Canada, and
the Gulf of Mexico.

\section{Neural Network, Data, and Training}
\label{sec:nn}

\subsection{Input Features and Training Data}
\label{sec:features}

\begin{figure}[H]
\centering
\includegraphics[width=0.85\textwidth]{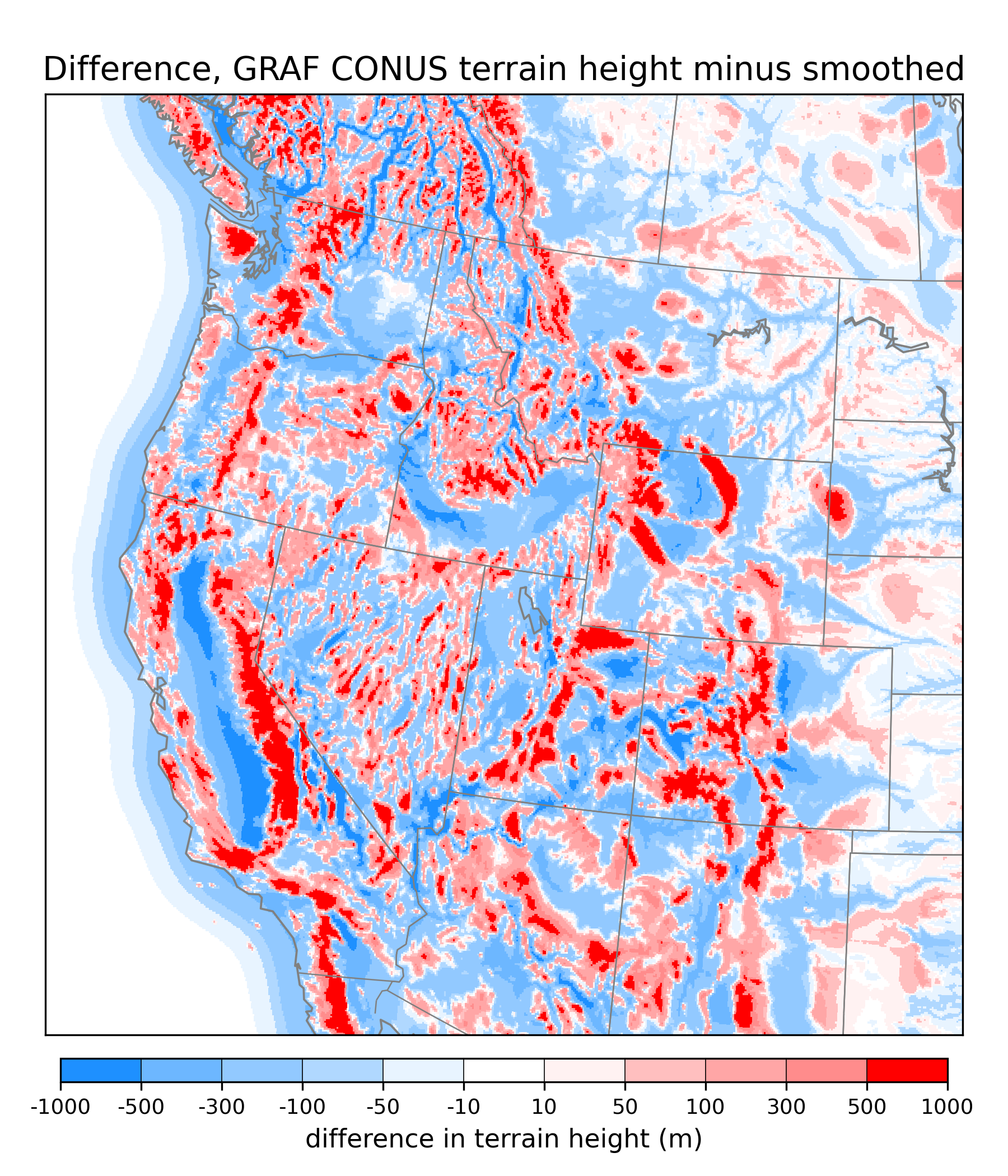}
\caption{Terrain height differences from a smoothed terrain representation shown 
	in the western US, used as a feature in the Attention ResUNet training.  The difference
	from smoothed terrain accentuates local elevation differences.}
\label{fig:terrain_diff}
\end{figure}

The Attention ResUNet model, described later, is trained 
on $96 \times 96$ pixel patches extracted from
the CONUS domain.  Each patch comprises seven input channels:

\begin{enumerate}[label=(\arabic*)]
  \item \textbf{GRAF precipitation forecast} ($g$): This is the raw
        model-predicted hourly accumulated precipitation (mm),
        transformed by a power function $g' = g^{p}$ with
        $p = 0.5$ (square root) to minimize training sensitivity to outliers 
        and variance stabilization.
        (Section~\ref{sec:power}).
  \item \textbf{Terrain elevation deviation} ($\Delta z$): This is the
        difference between local terrain height and a smoothed
        representation of terrain height provided by a Gaussian
        convolution with a length scale of 15 grid points
        (Figure~\ref{fig:terrain_diff}), which facilitates
        definition of local orographic variability.
  \item \textbf{GFS column-average relative humidity} ($R$):
        This provides large-scale moisture information from NOAA's Global Forecast
        System (GFS) model, interpolated to the GRAF grid.  Ideally this would
        be provided directly by the GRAF system, but we neglected to save
        a long time series of these.  Use of this was found to help distinguish between
        no probability of precipitation and small probability.
  \item \textbf{Precipitation--terrain interaction} ($g' \times
        \Delta z$): This was chosen to enhance potential terrain-enhanced precipitation
        processes (upslope enhancement, rain shadowing).
  \item \textbf{Precipitation--humidity interaction} ($g' \times
        R$): This is intended to help estimate the modulation of precipitation probability due
        to the large-scale moisture environment.
  \item \textbf{Terrain gradient, zonal} ($\partial z / \partial
        \lambda$): This is the longitudinal component of terrain slope, where terrain
        elevation was smoothed with a Gaussian kernel with a length scale of 15 grid points.
        This is intended to facilitate estimation of larger-scale terrain-induced precipitation.
  \item \textbf{Terrain gradient, meridional} ($\partial z /
        \partial \phi$): This is the latitudinal component of terrain slope, also smoothed.
\end{enumerate}

Note that interaction terms (4) and (5) are computed \emph{after}
the power transformation has been applied to the GRAF precipitation.

The choice of features reflected what data were readily available. At TWCo,
we have maintained an archive of hourly precipitation from GRAF dating back
several years, but we did not archive long time series of other variables that
might have become useful features, which might have included low-level winds
and relative humidity, static stability, and so forth.   To make up for the lack
of an archive of GRAF relative humidity, we substitute GFS humidity data.   
As of March 2026, TWCo started archiving more variables to facilitate only 
using features from the GRAF model in future studies.

\subsubsection{Power transformation}
\label{sec:power}

Precipitation amounts are strongly right-skewed.  Applying a power
transformation $g' = g^p$ with $p < 1$ compresses the range of
large values and stabilizes variance, which is beneficial because
the Gamma distribution's variance scales with the square of its
mean.  For Gamma-distributed data, the square root ($p = 0.5$) is
asymptotically the variance-stabilizing transformation
\citep{bartlett1936}.  The power transformation is applied to
GRAF precipitation \emph{before} computing the interaction
features and before normalization.

\subsubsection{Normalization}
\label{sec:normalization}

Each feature channel is independently normalized to $[0, 1]$ using
fixed reference maxima chosen to accommodate extreme values:
\begin{equation}
  \tilde{x}_k = \frac{x_k - x_k^{\min}}{x_k^{\max} - x_k^{\min}},
  \label{eq:normalization}
\end{equation}
where the reference values (Table~\ref{tab:normalization}) are
determined from domain knowledge and training-set statistics.
Using fixed maxima rather than sample-dependent standardization
ensures consistent normalization across training, validation, and
inference.

\begin{table}[ht]
\centering
\caption{Normalization reference values for each input channel
  when $p = 0.5$ (square root transformation).}
\label{tab:normalization}
\begin{tabular}{lcc}
\toprule
Channel & $x_k^{\min}$ & $x_k^{\max}$ \\
\midrule
GRAF precipitation ($g^{0.5}$) & 0 & $75^{0.5} \approx 8.66$ \\
Terrain deviation ($\Delta z$) & data min & 2500 \\
GFS relative humidity ($R$) & data min & 100 \\
$g^{0.5} \times \Delta z$ & data min & $75^{0.5} \times 2500$ \\
$g^{0.5} \times R$ & data min & $75^{0.5} \times 100$ \\
$\partial z / \partial \lambda$ & data min & 0.02 \\
$\partial z / \partial \phi$ & data min & 0.02 \\
\bottomrule
\end{tabular}
\end{table}

\subsubsection{MRMS Target Data}
\label{sec:mrms}

The training targets are derived from the Multi-Radar Multi-Sensor
(MRMS) quantitative precipitation estimation system \citep{zhang2016},
which merges data from the U.S.\ Weather Surveillance Radar--1988 Doppler (WSR-88D) network, gauge
reports, and satellite retrievals to produce high-resolution,
near-real-time precipitation analyses on a 0.01$^{\circ}$
($\approx$1\,km) grid over the CONUS.

\paragraph{Remapping to the GRAF grid.}
Hourly MRMS precipitation analyses are read from files of the Stage\,II
multi-sensor pass-2 product.  A collocated MRMS
data-quality file provides a per-pixel quality index on a
$[0,\,1]$ scale encoding sensor coverage and retrieval confidence.
Both fields are remapped from the native MRMS 0.01$^{\circ}$
($\approx$1\,km) grid to the GRAF CONUS grid ($\approx$4.25\,km
spacing, hereafter "4 km") using nearest-neighbor interpolation.
Nearest-neighbor interpolation was chosen over other alternatives
such as performing the training at 1 km, bilinear interpolation to 4 km,
or a conservative upscaling to the 4-km grid.
Keeping the 1-km grid spacing would have
increased training time for this preliminary study.
Were the analyses upscaled to 4 km rather than
sampling, the target data would not represent the local variability that we expect
customers want in their probability forecasts.

\paragraph{Quality masking.}
MRMS grid points with a quality index $\leq 0.01$ are treated as unreliable and are
not used in the training. During patch extraction, potential 
tiles in which more than 50\% of pixels fail this quality threshold
are discarded from consideration as a training sample.  Within retained tiles, 
individual pixels with the quality index $\leq 0.01$ 
receive a target value of $-1$; the loss function treats this as a
mask and excludes such pixels from gradient computation.  This
strategy preserves spatial coverage in partially contaminated tiles
while preventing the network from learning from poor-quality radar
retrievals.

The quality threshold of 0.01 used in training is admittedly quite non-restrictive.  
Had we used a higher quality threshold such as 0.5, say, the target data would
be more trustworthy.   However, this would come at the expense of providing
much less training data to the algorithm in the western US.  When
quality is low due to problems like radar beam blockage, the target MRMS analysis
data is more reflective of the model background forecast provided by the
NOAA High-Resolution Rapid Refresh \citep[HRRR;][]{benjamin2016, dowell2022}.
Because of this, in mountainous regions the precipitation analyses may reflect
HRRR model bias, making a biased target for training.   Despite the low threshold
for training, for model verification, we will limit the acceptable samples to those
with quality $>$ 0.5.

\subsubsection{Data sampling}
\label{sec:sampling}

Training patches are drawn from four 60-day date windows that
together provide both recency and seasonal variety:
(1)~the 60 days immediately preceding the initialization date
(offset slightly so the valid time falls outside the window);
(2)~a 60-day window spanning approximately 10--12 months before
the initialization date; (3)~a 60-day window spanning approximately
12--14 months before the initialization date; and (4)~a 60-day
window spanning approximately 22--24 months before the
initialization date.   Dates within each window are sampled with 
initial conditions at 6-hourly intervals.  Separate trainings are conducted
for every month of the year.

Given the lead-time dependence of skill, a separate training is performed 
every 3 h of forecast lead time from +3 h to +48 h.   To minimize training
time, the training for 6-h forecasts, conducted after the training of 3-h 
forecasts, start from the weight estimates from the +3 h training rather
than from random initial weights.

Patch data are randomized in several ways to prevent over-learning
to terrain features.  First, the center locations of the patches on any given day being
sampled is random.
For each date, the CONUS domain is tiled with a non-overlapping
$96\times96$ grid.  The origin of the grid is shifted by a
date-seeded random offset $(s_y, s_x) \in [0, 96)^2$, so that
terrain features appear at different patch-local positions across
training days, reducing the risk of over-learning fixed terrain
patterns.  Tiles whose domain-average MRMS quality fraction of
bad pixels exceeds 50\% are discarded outright; individual bad
pixels within retained tiles are masked during training.

Tiles are partitioned into \emph{wet} (patch maximum GRAF
$\geq 0.5$\,mm) and \emph{dry} subsets.  Wet tiles are sampled
with probability proportional to (patch-mean precipitation)$^{1.5}$,
biasing the training set toward regions of active precipitation.
The number of wet tiles per day, $n_{\text{wet}}$, depends on
domain wetness:  For active days, defined as where the domain mean $> 0.15$\,mm, 
35 wet samples are chosen.   For moderate days, where the domain mean is between
$0.10$--$0.15$\,mm, 25 samples are chosen. Finally, for dry days,
where the domain mean $< 0.10$\,mm, 10 samples are chosen.
An additional $n_{\text{dry}} \in \{5, 6, 7, 8\}$ dry tiles are
selected uniformly at random.  

\subsubsection{Data augmentation}
\label{sec:augmentation}

Training patches undergo random geometric augmentation; there is a 50\%
probability of applying a horizontal flip, whereby the image is
reflected left-to-right and the zonal terrain gradient
channel ($\partial z / \partial \lambda$) is multiplied by $-1$.
There is also a 50\% probability of reflecting
top-to-bottom, and in this case the meridional terrain gradient
channel ($\partial z / \partial \phi$) is multiplied by $-1$.
Multiplying the gradient channels by $-1$ preserves physical
consistency: under a left-right reflection a west-facing slope
becomes an east-facing slope, so the sign of the zonal gradient
must be reversed, and analogously for the meridional gradient under
a top-bottom reflection.

Random geometric augmentation artificially expands the effective
size of the training set, exposing the network to a broader variety
of spatial configurations.  It is particularly useful here because
CONUS terrain and precipitation patterns are not
left-right or top-bottom symmetric; without augmentation the
network could over-fit to the directional biases present in the
training patches.

\subsection{Network Architecture}
\label{sec:architecture}

\begin{figure}[H]
\centering
\includegraphics[width=\textwidth]{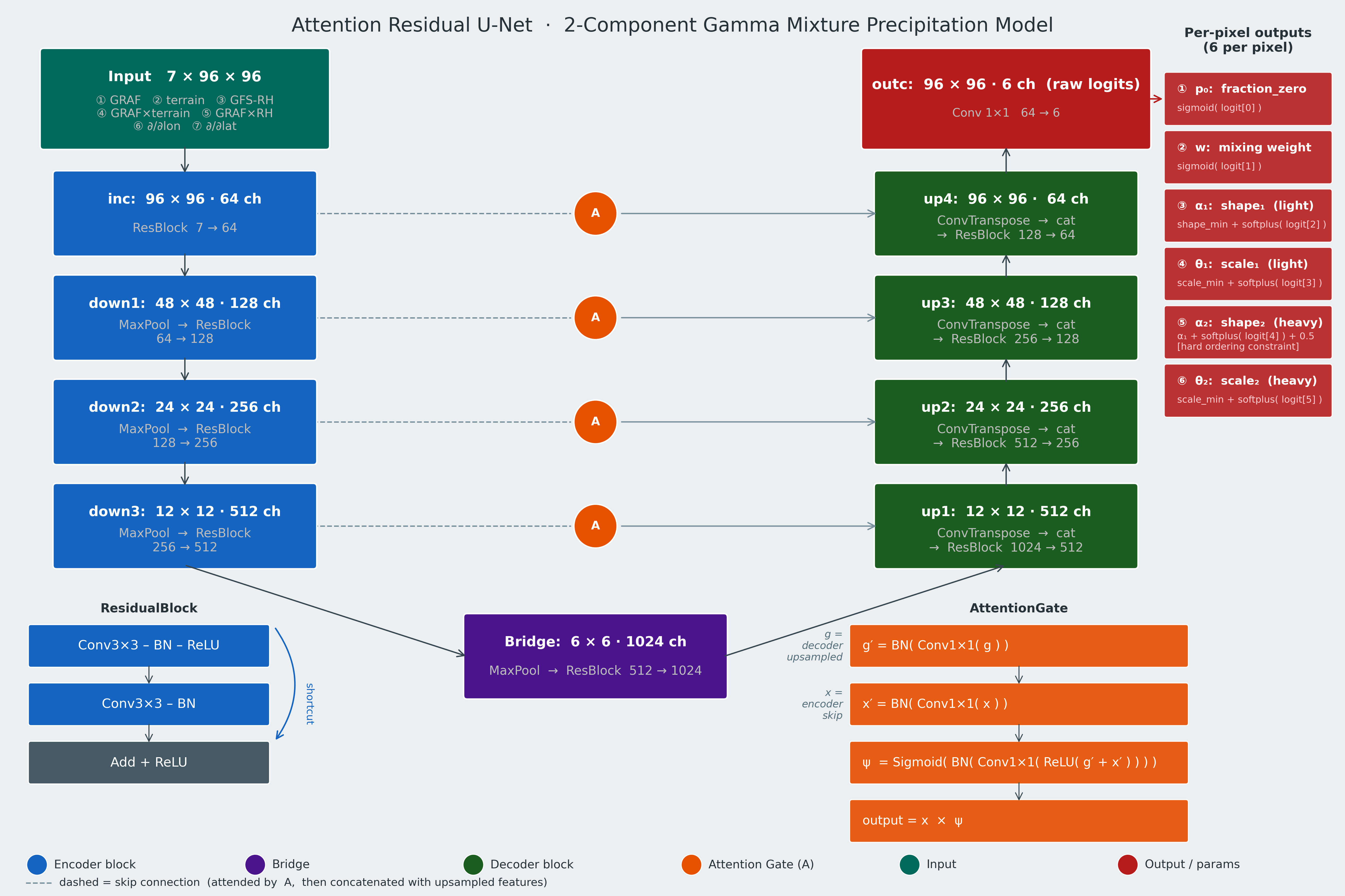}
\caption{Architecture of the Attention Residual U-Net for two-component
  Gamma mixture precipitation forecasting.
  \emph{Left (encoder):} The input is a $7 \times 96 \times 96$
  patch comprising seven channels: GRAF precipitation
  (power-transformed), terrain elevation deviation, GFS column-average
  relative humidity, precipitation$\times$terrain and
  precipitation$\times$humidity interaction terms, and zonal and
  meridional terrain gradients.  An initial Residual Block (inc, 64
  channels) is followed by four $2\times2$ max-pooling
  downsampling stages that halve the spatial dimensions and double the
  channel count at each level (128, 256, 512 channels at
  $48\times48$, $24\times24$, and $12\times12$, respectively).
  The ResidualBlock inset (bottom left) shows the structure: two
  Conv$3\times3$--BN--ReLU layers with a shortcut connection to
  mitigate gradient vanishing.
  \emph{Center (bridge):} A Residual Block at the coarsest resolution
  ($6\times6$, 1024 channels) integrates broad spatial context
  spanning the full $\approx$384\,km extent of the patch.
  \emph{Right (decoder):} Four upsampling stages apply $2\times2$
  transposed convolutions, progressively recovering spatial detail.
  At each stage an Attention Gate (orange diamond, labelled
  \textbf{A}) computes a spatial weighting map $\nu$ that
  re-weights the corresponding encoder skip-connection features
  before concatenation with the upsampled decoder features; the
  concatenated tensor then passes through a Residual Block.  The
  AttentionGate inset (bottom right) illustrates how decoder query
  features $g$ and encoder key features $x$ are combined via learned
  $1\times1$ projections, summed, ReLU-activated, projected to a
  scalar channel, and passed through a sigmoid to produce the
  element-wise gate $\nu \in (0,1)$.
  \emph{Output:} A final $1\times1$ convolution maps 64 channels to
  six raw logits.  The per-pixel output panel (red, far right) shows
  the six distributional parameters after activation: $p_0$
  (zero-precipitation probability, sigmoid), $w$ (conditional mixture
  weight, sigmoid), $\alpha_1$ (light-component shape, softplus),
  $\theta_1$ (light-component scale, softplus), $\alpha_2$
  (heavy-component shape, softplus with $\alpha_2>\alpha_1$ enforced
  post-forward), and $\theta_2$ (heavy-component scale, softplus).}
\label{fig:arch}
\end{figure}

The model is an Attention Residual U-Net (Fig.~\ref{fig:arch}),
combining the U-shaped encoder--decoder architecture of
\citet{ronneberger2015} with residual connections \citep{he2016} and
attention gates \citep{oktay2018}.  The network maps a
$7 \times 96 \times 96$ input tensor to a $6 \times 96 \times 96$
output tensor of six raw real-valued logits per pixel, which are
subsequently mapped through sigmoid and softplus activations to the
constrained distributional parameters of the zero-inflated
two-component Gamma mixture (Section~\ref{sec:distribution}).

\subsubsection{Residual blocks}
\label{sec:resblock}

Each encoder and decoder stage uses a \emph{Residual Block}
\citep{he2016}, consisting of two $3 \times 3$ convolutional
layers with batch normalization \citep[BN;][]{ioffe2015} and
rectified linear unit (ReLU) activation, plus an identity shortcut:
\begin{equation}
  \vect{y} = \text{ReLU}\!\Big(
    \underbrace{
      \text{BN}\big(\text{Conv}_{3\times3}\big(
        \text{ReLU}\big(\text{BN}(\text{Conv}_{3\times3}(\vect{x}))\big)
      \big)\big)
    }_{\text{main path}}
    + \underbrace{W_s \vect{x}}_{\text{shortcut}}
  \Big),
  \label{eq:resblock}
\end{equation}
where $W_s$ is a $1 \times 1$ convolution applied when the number
of input and output channels differ, and the identity otherwise.
Residual connections mitigate gradient vanishing in deeper networks
\citep{he2016}.  In the context of precipitation post-processing
they offer an additional advantage: the residual formulation biases
the network toward identity mappings, so when the GRAF forecast is
already accurate in a region the network learns only a small
correction rather than reconstructing the full feature
representation from scratch.  This is particularly valuable in
encoder stages that process predominantly dry patches, where the
optimal transformation is close to the identity.  Residual blocks
also stabilize optimization when the NLL loss exhibits the large
dynamic range characteristic of mixed dry/wet batches, since
gradient magnitudes are bounded by the shortcut path even when
the main-path activations are small.

\subsubsection{Encoder}
\label{sec:encoder}

The encoder's role is to progressively compress the spatial domain
while expanding the channel dimension, building a hierarchy of
representations at successively coarser scales.  Shallow encoder
stages retain fine spatial detail and respond to local precipitation
gradients and terrain edges; deeper stages integrate information
across larger areas, capturing mesoscale and synoptic-scale patterns.
This hierarchical feature extraction is what allows the network to
relate broad moisture environments to local precipitation at the
end of the decoder.

The encoder comprises four downsampling stages.  Each stage applies
$2 \times 2$ max-pooling followed by a residual block, doubling
the channel count:

\begin{center}
\begin{tabular}{lccc}
\toprule
Stage & Input channels & Output channels & Spatial size \\
\midrule
Initial & 7 & 64 & $96 \times 96$ \\
Down 1 & 64 & 128 & $48 \times 48$ \\
Down 2 & 128 & 256 & $24 \times 24$ \\
Down 3 & 256 & 512 & $12 \times 12$ \\
\bottomrule
\end{tabular}
\end{center}

\subsubsection{Bridge}
\label{sec:bridge}

The bridge operates at the coarsest resolution ($6 \times 6$) with 512 input channels,
expanding to 1024 channels, capturing the largest-scale spatial context.

\subsubsection{Attention gates}
\label{sec:attention}

Standard U-Net skip connections \citep{ronneberger2015} concatenate
encoder features with decoder features at matching resolutions.
However, not all encoder features are equally informative at every
spatial location.  Attention gates \citep{oktay2018} learn to
suppress irrelevant regions and highlight salient features.

Given decoder features $\vect{g}$ (the ``query'') and encoder
features $\vect{x}$ (the ``key/value''), the attention gate
computes:
\begin{align}
  \vect{q}_g &= \text{BN}(W_g \vect{g}), \label{eq:att_g} \\
  \vect{q}_x &= \text{BN}(W_x \vect{x}), \label{eq:att_x} \\
  \nu &= \sigmoid\!\big(\text{BN}(\psi \cdot \text{ReLU}(
    \vect{q}_g + \vect{q}_x))\big), \label{eq:att_nu} \\
  \hat{\vect{x}} &= \nu \odot \vect{x}, \label{eq:att_out}
\end{align}
where $W_g$, $W_x$, and $\psi$ are learnable $1 \times 1$
convolutions, $\sigmoid(\cdot)$ denotes the sigmoid activation, and $\odot$
denotes element-wise multiplication.  The intermediate channel
dimension ($F_{\text{int}}$) is half the decoder channel count.

The incorporation of attention gates provides several advantages specific to the spatial
post-processing of precipitation fields; the first is spatial selectivity.
Precipitation is highly localized, and large portions of any domain
patch may be dry at any given time.  The attention map $\nu$ learns to
suppress encoder features from dry, featureless regions,
preventing them from diluting the decoder's representation
of active precipitation areas or areas with complex terrain (not shown).  
In practice the learned $\nu$ maps cluster near zero over dry regions and near
unity over precipitating or orographically active regions,
consistent with this intended role.  Second, at the coarsest decoder
scale ($12\times12$, corresponding to $\approx$384\,km),
the attention gate draws on broad synoptic-scale encoder
context such as the moisture plume associated with an
atmospheric river to guide upsampling.  At the finest
scale ($96\times96$, $\approx$4\,km), it refines that
signal using fine-scale terrain gradients.  This multi-scale
gating lets the network modulate which features are
forwarded at each resolution stage independently.

A third advantage of the attention mechanism is that in
mountainous patches, terrain gradients strongly govern precipitation through
upslope enhancement and rain-shadow effects; in flat
patches, large-scale moisture is more decisive.  Rather
than concatenating all encoder features with equal weight,
the attention gate can up-weight terrain-gradient channels
when orographic forcing is evident in the GRAF forecast
and suppress them when precipitation is large-scale and
independent of local slope.  Finally, without attention,
irrelevant encoder activations---e.g., terrain information
from a dry, cloudless corner of a patch---pass unchanged
through the skip connection and must be suppressed by
subsequent decoder convolutions.  The attention gate
zeros those contributions at the source, simplifying
the decoder's learning task.

\subsubsection{Decoder}
\label{sec:decoder}

The decoder's purpose is to progressively recover the spatial
detail that was compressed by the encoder's max-pooling stages.
The decoder mirrors the encoder with four upsampling stages.
Each stage applies a $2 \times 2$ transposed convolution (halving
channels), passes the encoder features through an attention gate,
concatenates the attended encoder features with the upsampled
decoder features, and processes the result through a residual block.
At coarse scales (Up\,1--2) it integrates the synoptic-scale context
encoded at the bridge with mesoscale encoder features, establishing
the broad structure of the predicted probability field.  At finer
scales (Up\,3--4) it incorporates terrain-gradient and local
precipitation features from the shallow encoder layers, enabling
the output to reflect kilometer-scale orographic modulation.
Because attended skip connections deliver encoder features directly
to the matching decoder stage, fine spatial detail that would
otherwise be irreversibly lost during max-pooling is recovered
without requiring the decoder to reconstruct it from the compressed
bridge representation alone.

\begin{center}
\begin{tabular}{lcccc}
\toprule
Stage & Input ch.\ (up) & Attn.\ $F_{\text{int}}$ &
  Concat ch. & Output ch. \\
\midrule
Up 1 & 1024 $\to$ 512 & 256 & 1024 & 512 \\
Up 2 & 512 $\to$ 256 & 128 & 512 & 256 \\
Up 3 & 256 $\to$ 128 & 64 & 256 & 128 \\
Up 4 & 128 $\to$ 64 & 32 & 128 & 64 \\
\bottomrule
\end{tabular}
\end{center}

\subsubsection{Output layer}
\label{sec:output_layer}

The final $1 \times 1$ convolution maps 64 decoder channels to
$C = 6$ output channels, one for each unconstrained parameter of
the zero-inflated two-Gamma mixture distribution
(Section~\ref{sec:distribution}).

\subsection{Predictive Distribution}
\label{sec:distribution}

At each pixel, the model predicts a \emph{zero-inflated mixture of
two Gamma distributions}.  The distribution is hierarchical:
\begin{enumerate}[label=(\roman*)]
  \item with probability $p_0$, precipitation is exactly zero;
  \item with probability $1 - p_0$, precipitation is positive and
        drawn from a two-component Gamma mixture:
        \begin{itemize}
          \item light-precipitation component
                ($\text{Gamma}(\alpha_1, \theta_1)$), conditional weight $w$,
          \item heavy-precipitation component
                ($\text{Gamma}(\alpha_2, \theta_2)$), conditional weight $1-w$.
        \end{itemize}
\end{enumerate}
The weights $w$ and $1-w$ are the \emph{conditional} mixture weights given
positive precipitation, so they sum to unity.

\begin{figure}[H]
\centering
\includegraphics[width=\textwidth]{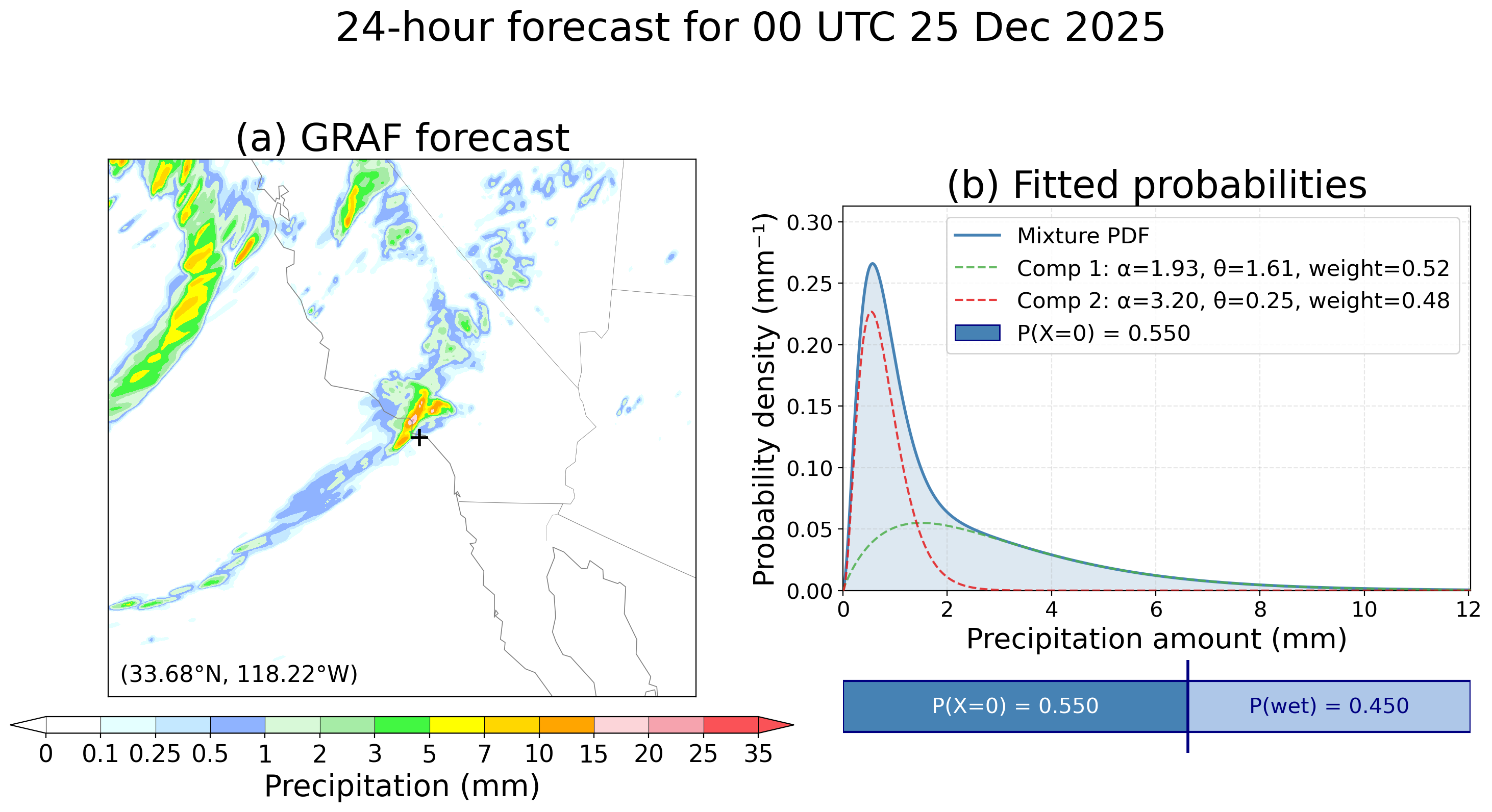}
\caption{Example predictive density of the zero-inflated mixture
  of two Gamma distributions at a single grid point.  The point
  mass at $y=0$ (height $p_0$) represents the probability of no
  precipitation.  The continuous density for $y>0$ is a weighted
  mixture of a light-precipitation Gamma component
  ($\text{Gamma}(\alpha_1,\theta_1)$, dashed) and a
  heavy-precipitation component
  ($\text{Gamma}(\alpha_2,\theta_2)$, dash-dot), scaled by
  $(1-p_0)$; their sum is the full conditional density (solid).}
\label{fig:graf_pdf}
\end{figure}

The network outputs six unconstrained real values
$(z_0, z_1, z_2, z_3, z_4, z_5)$ per pixel, which are
transformed to satisfy the distributional constraints.
The zero-precipitation probability $p_0$ and the conditional
mixture weight $w$ are obtained independently by applying the
sigmoid function:
\begin{align}
  p_0 &= \sigmoid(z_0) = \frac{1}{1 + e^{-z_0}},
    \label{eq:sigmoid_p0} \\
  w   &= \sigmoid(z_1) = \frac{1}{1 + e^{-z_1}}.
    \label{eq:sigmoid_w}
\end{align}
This guarantees $p_0, w \in (0, 1)$, so $p_0$ is a valid
probability of zero precipitation and $w$ is a valid conditional
mixing weight.
The Gamma shape and scale parameters must be strictly positive.
The first component uses softplus with minimum bounds:
\begin{align}
  \alpha_1 &= \alpha_{\min} + \softplus(z_2)
    = \alpha_{\min} + \log(1 + e^{z_2}),
    \label{eq:shape1} \\
  \theta_1 &= \theta_{\min} + \softplus(z_3)
    = \theta_{\min} + \log(1 + e^{z_3}),
    \label{eq:scale1}
\end{align}
where $\alpha_{\min}$ and $\theta_{\min}$ are small positive
constants (Section~\ref{sec:initialization}) that prevent
numerical instability in the log-gamma and logarithm computations.
To prevent label switching (where the two Gamma components swap
roles during training), the second shape parameter is constrained
to exceed the first:
\begin{equation}
  \alpha_2 = \alpha_1 + \softplus(z_4) + \delta_{\min},
  \label{eq:shape2_constraint}
\end{equation}
where $\delta_{\min} = 0.5$ enforces a minimum separation between
components.  This guarantees $\alpha_2 > \alpha_1$ at all times,
ensuring that component~1 always represents lighter precipitation.
The second scale parameter is unconstrained relative to $\theta_1$:
\begin{equation}
  \theta_2 = \theta_{\min} + \softplus(z_5).
  \label{eq:scale2}
\end{equation}
Both $\alpha_j$ and $\theta_j$ are additionally clamped to maximum
values ($\alpha_{\max}$, $\theta_{\max}$) to prevent overflow.

In summary, the six constrained parameters are:
\begin{equation}
  \underbrace{p_0}_{\sigmoid(z_0)},\quad
  \underbrace{w}_{\sigmoid(z_1)},\quad
  \underbrace{(\alpha_1, \theta_1)}_{\text{softplus of } (z_2, z_3)},\quad
  \underbrace{\alpha_2}_{\alpha_1 + \softplus(z_4) + \delta_{\min}},\quad
  \underbrace{\theta_2}_{\text{softplus}(z_5)}.
  \label{eq:all_params}
\end{equation}

\subsubsection{Probability density function}
\label{sec:pdf}

Let $Y$ denote the precipitation amount at a pixel.  The predictive
distribution is:
\begin{equation}
  F_Y(y) =
  \begin{cases}
    p_0, & y = 0, \\[4pt]
    p_0 + (1 - p_0)\!\left[w \, F_{\Gamma}(y; \alpha_1, \theta_1)
        + (1-w) \, F_{\Gamma}(y; \alpha_2, \theta_2)\right],
    & y > 0,
  \end{cases}
  \label{eq:mixture_cdf}
\end{equation}
where $F_{\Gamma}(y; \alpha, \theta)$ is the CDF of the
$\text{Gamma}(\alpha, \theta)$ distribution.  The factor $(1-p_0)$
converts the conditional mixture weights $(w,\,1-w)$ to
unconditional probabilities.

The density for $y > 0$ is:
\begin{equation}
  f_Y(y) = (1 - p_0)\!\left[w \, f_{\Gamma}(y; \alpha_1, \theta_1)
          + (1-w) \, f_{\Gamma}(y; \alpha_2, \theta_2)\right],
  \label{eq:mixture_pdf}
\end{equation}
with the Gamma PDF:
\begin{equation}
  f_{\Gamma}(y; \alpha, \theta)
    = \frac{1}{\Gamma(\alpha)\,\theta^{\alpha}}
      \, y^{\alpha - 1} \, e^{-y/\theta},
  \quad y > 0.
  \label{eq:gamma_pdf}
\end{equation}

\subsubsection{Exceedance probabilities}
\label{sec:exceedance}

For any threshold $\tau > 0$, the exceedance probability follows
from~\eqref{eq:mixture_cdf}:
\begin{equation}
  P(Y > \tau) = (1 - p_0)\Big\{
    w \big[1 - F_{\Gamma}(\tau; \alpha_1, \theta_1)\big]
    + (1-w) \big[1 - F_{\Gamma}(\tau; \alpha_2, \theta_2)\big]
  \Big\}.
  \label{eq:exceedance}
\end{equation}
Standard operational thresholds include 0.25, 1.0, 2.5, 5.0,
10.0, and 25.0\,mm.  The Gamma CDF is evaluated using the
regularized incomplete gamma function, which is available in
standard numerical libraries.

\subsection{Loss Function}
\label{sec:loss}

The model is trained by minimizing the \emph{negative
log-likelihood} (NLL), which is a strictly proper scoring rule
\citep{gneiting2007}.  The NLL was preferred over the Continuous Ranked Probability Score
(CRPS), the other common strictly proper scoring rule used in probabilistic
post-processing.  For the zero-inflated two-component Gamma mixture,
the CRPS has no closed-form expression and must be approximated by
numerical integration or Monte Carlo sampling of the mixture CDF at
many threshold values---an expensive inner loop for every pixel in
every mini-batch.  In contrast, the Gamma log-PDF has an analytic
form (leveraging PyTorch's \texttt{torch.lgamma} for gradient support), so
the NLL can be evaluated exactly in a single vectorized pass.
Minimizing the NLL is also equivalent to maximum-likelihood
estimation of the distributional parameters, providing a direct
link to the parametric model being trained.

For a single pixel with observation $y$ and predicted parameters $(p_0, w, \alpha_1, \theta_1, \alpha_2, \theta_2)$, the NLL is:

\paragraph{Case 1: $y = 0$ (dry observation).}
\begin{equation}
  \text{NLL} = -\log p_0.
  \label{eq:nll_zero}
\end{equation}

\paragraph{Case 2: $y > 0$ (wet observation).}
The NLL decomposes into a zero-inflation term and a mixture term:
\begin{equation}
  \text{NLL} = -\log(1 - p_0)
    - \log\!\big[
        w \, f_{\Gamma}(y; \alpha_1, \theta_1)
        + (1-w) \, f_{\Gamma}(y; \alpha_2, \theta_2)
      \big],
  \label{eq:nll_positive}
\end{equation}
where the first term penalizes assigning too much zero-probability
mass when precipitation is observed, and the Gamma log-PDF is:
\begin{equation}
  \log f_{\Gamma}(y; \alpha, \theta)
    = (\alpha - 1) \log y - \frac{y}{\theta}
      - \log \Gamma(\alpha) - \alpha \log \theta.
  \label{eq:gamma_logpdf}
\end{equation}

The mixture log-likelihood term is evaluated using the
log-sum-exp trick for numerical stability:
\begin{equation}
  -\log\!\bigl(w f_1 + (1-w) f_2\bigr)
  = -\log\!\big(e^{\log w + \ell_1} + e^{\log(1-w) + \ell_2}\big),
  \label{eq:logsumexp}
\end{equation}
where $\ell_j = \log f_{\Gamma}(y; \alpha_j, \theta_j)$.  The
$\log\Gamma(\alpha)$ term is computed via \texttt{torch.lgamma},
which provides full gradient support.

\paragraph{Masking.}
Pixels with poor MRMS radar quality (quality index
$\leq 0.01$) are assigned a target value of $-1$ and excluded
from the loss computation.

\paragraph{Batch loss.}
The overall loss is the mean NLL over all valid pixels in the
mini-batch:
\begin{equation}
  \mathcal{L} = \frac{1}{|\mathcal{V}|}
    \sum_{i \in \mathcal{V}} \text{NLL}_i,
  \label{eq:batch_loss}
\end{equation}
where $\mathcal{V}$ is the set of valid (non-masked) pixels.

\subsection{Climatological Initialization}
\label{sec:initialization}

Neural network outputs are initially near zero, which after
sigmoid/softplus transformations would produce poorly calibrated
starting predictions.  We initialize the output layer's biases
using climatological statistics derived by an Expectation--Maximization
(EM) algorithm applied to wet pixels from the training set.
This initialization places the network's predictions close to the
marginal climatological distribution from the first forward pass,
avoiding the large NLL values that arise when the predicted $p_0$
or the Gamma parameters are far from climatology.  The result is
noticeably faster convergence: in practice, early-stopping patience
is triggered several epochs sooner than with default (near-zero)
initialization.

\begin{enumerate}
  \item \textbf{Fraction of zeros.}  Compute $\hat{p}_0$, the
        fraction of valid training pixels with zero precipitation.
        Set $b_0 = \logit(\hat{p}_0) \equiv \log\!\bigl(\hat{p}_0 /
        (1 - \hat{p}_0)\bigr)$.

  \item \textbf{Two-component Gamma mixture via EM.}  Up to 50\,000
        wet pixels are sub-sampled from the training set and a
        two-component Gamma mixture is fitted by the EM algorithm
        (up to 500 iterations, initialized by method of moments).
        The EM returns mixture weight $\hat{w}$, shape parameters
        $(\hat{\alpha}_1, \hat{\alpha}_2)$, and scale parameters
        $(\hat{\theta}_1, \hat{\theta}_2)$, with components ordered
        so that $\hat{\alpha}_1 < \hat{\alpha}_2$ (light before heavy).

  \item \textbf{Bias initialization.}  The output layer biases are set
        via the inverse sigmoid and inverse softplus:
        \begin{align}
          b_0 &= \logit(\hat{p}_0), \\
          b_1 &= \logit(\hat{w}), \\
          b_2 &= \log\!\bigl(e^{\hat{\alpha}_1 - \alpha_{\min}} - 1\bigr), \\
          b_3 &= \log\!\bigl(e^{\hat{\theta}_1 - \theta_{\min}} - 1\bigr), \\
          b_4 &= \log\!\bigl(e^{\hat{\alpha}_2 - \hat{\alpha}_1 - \delta_{\min}} - 1\bigr), \\
          b_5 &= \log\!\bigl(e^{\hat{\theta}_2 - \theta_{\min}} - 1\bigr),
        \end{align}
        where the argument of the log is clamped to be positive before
        evaluation.  The output layer weights are initialized with a
        small Xavier uniform random initialization.

  \item \textbf{Minimum bounds.}  $\alpha_{\min} = 0.3$ (fixed);
        $\theta_{\min} = 0.01$ (fixed); $\delta_{\min} = 0.5$
        (minimum shape separation enforced by the hard ordering
        constraint, \eqref{eq:shape2_constraint}).
\end{enumerate}

\subsection{Training}
\label{sec:training}

\subsubsection{Optimization}
\label{sec:optimization}

Training uses the Adam optimizer \citep{kingma2015} with a base learning
rate of $\eta = 7 \times 10^{-4}$, processing batches of 128 samples
loaded directly onto the GPU without gradient accumulation.  The learning
rate is adapted by a PyTorch \texttt{ReduceLROnPlateau} scheduler
(reduction factor 0.7, patience 2 epochs) monitoring validation loss.  Training
halts when validation loss fails to improve for 7 consecutive epochs, up
to a maximum of 40 epochs.  Gradients are clipped to a maximum norm of
1.0 (when the power transformation is applied) or 2.0 (without) to
prevent explosion from the $\log\Gamma$ and logarithm operations.  On
CUDA (Compute Unified Device Architecture) devices, the forward pass uses
bfloat16 precision, which retains the dynamic range of float32, while the
NLL loss is evaluated in float32 to avoid numerical instability in the
$\log\Gamma$ and logarithm operations.
When no checkpoint exists for the target lead time, training warm-starts
from the nearest available lead time's weights.

The hyperparameter values above reflect an empirical balance
between training speed and forecast accuracy.  The base learning
rate of $7\times10^{-4}$ was found to converge reliably across lead
times without requiring a warm-up schedule; lower values slowed
convergence noticeably while higher values occasionally destabilized
the NLL loss during early epochs.  The \texttt{ReduceLROnPlateau}
scheduler's patience of~2 tolerates transient validation-loss
fluctuations without prematurely decaying the learning rate, while
the early-stopping patience of~7 prevents over-fitting when the
validation loss plateaus.  The batch size of~128 was the largest
that fit comfortably in GPU memory while still providing sufficient
gradient diversity from the mixed wet/dry training patches; smaller
batches increased variance in the gradient estimate and slowed
wall-clock convergence.

\subsubsection{Numerical stability}
\label{sec:stability}

Several safeguards prevent numerical issues during training:
Gamma parameters are clamped: $\alpha \in [\alpha_{\min},
\alpha_{\max}]$, $\theta \in [\theta_{\min}, \theta_{\max}]$.
Observations and mixture weights are clamped away from
zero before taking logarithms: $y_{\text{safe}} = \max(y, \epsilon)$.
The per-pixel NLL is capped at $\text{NLL}_{\max}$ to
prevent a single outlier from dominating the batch loss.
Stability parameters are automatically adjusted based on
the power transformation setting (Table~\ref{tab:stability}).
These safeguards are necessary because the two-component Gamma
mixture NLL involves several operations that are numerically
hazardous in single-precision arithmetic.  The $\log\Gamma(\alpha)$
term diverges for large shape parameters, and $\log y$ diverges
as $y\to 0$.  Without clamping and the log-sum-exp stabilization
of~\eqref{eq:logsumexp}, a small number of pixels per batch---
typically either very heavy convective precipitation or near-zero
drizzle---can produce NLL values orders of magnitude larger than
the batch mean, destabilizing the optimizer.  The per-pixel cap
$\text{NLL}_{\max}$ bounds the influence of such outliers during
early training when model parameters are far from their optimum.
The power transformation ($p=0.5$) compresses the range of
precipitation values, so tighter clamp bounds and a lower NLL cap
suffice; the raw ($p=1$) settings are more permissive to accommodate
the wider dynamic range of untransformed heavy-precipitation events.

\begin{table}[ht]
\centering
\caption{Stability parameters as a function of the power
  transformation exponent $p$.}
\label{tab:stability}
\begin{tabular}{lcc}
\toprule
Parameter & $p < 1$ (transformed) & $p = 1$ (raw) \\
\midrule
$\epsilon$ & $10^{-4}$ & $10^{-6}$ \\
$\alpha_{\min}$ & 0.3 & 0.3 \\
$\alpha_{\max}$ & 100 & 200 \\
$\theta_{\min}$ & 0.01 & 0.01 \\
$\theta_{\max}$ & 100 & 200 \\
$\text{NLL}_{\max}$ & 100 & 200 \\
Gradient clip norm & 1.0 & 2.0 \\
\bottomrule
\end{tabular}
\end{table}

\subsection{Inference}
\label{sec:inference}

Since trainings were available only every three hours from lead times
of +3 to +48 h, there were no trainings for many hours, such as hour one 
of the forecast.   The 3-hour training weights whose lead time was nearest 
to that of the lead time of interest were used, e.g., 3-hour trainings were used 
for forecasts of lead times +1 to +4 h.

The trained Attention Residual U-Net is fully convolutional---it contains no
linear or flatten layers, relying entirely on convolutional, pooling,
and transposed-convolutional operations.  This means the network can
accept inputs of arbitrary spatial size at inference time, provided
both spatial dimensions are divisible by $2^4 = 16$ (the product of
the four max-pooling stages in the encoder). Using a domain-wide 
approach is simpler and faster than patch-based inference: it
requires no patch loop, no overlap management, and no weighted blending,
and it eliminates seam artifacts that can occur at patch boundaries.

For full-domain inference, the CONUS feature tensor
($7 \times 1308 \times 1524$) is padded with edge-replication to the
nearest multiple of 16 in each spatial dimension
($7 \times 1312 \times 1536$) and passed through the model in a single
forward pass, producing a $6 \times 1312 \times 1536$ output of raw
logits.  The output is cropped back to the native $1308 \times 1524$
domain.  The outermost 16 rows and columns of the output are masked as
missing, as they are influenced by the replicate-padded region and lack
real-data context outside the domain boundary.

After the single forward pass, the six raw logit fields are transformed to
constrained distributional parameters (sigmoid for $p_0$ and $w$;
softplus with the hard ordering constraint for the Gamma shape and scale
parameters), and exceedance probabilities are computed for each threshold
$\tau$ using~\eqref{eq:exceedance}.

\section{Reference Standard Probability Forecasts}
\label{sec:reference}

To establish a meaningful baseline for evaluating the Attention
Residual U-Net, we construct reference probabilistic forecasts
directly from the raw GRAF precipitation field by applying
Gaussian spatial smoothing.  The procedure follows the
``neighborhood probability'' approach \citep[e.g.,][]{theis2005, schwartz2015},
in which deterministic model output is convolved with a Gaussian
kernel to produce a probabilistic forecast.

For a given precipitation threshold $\tau$ and forecast date, a
binary exceedance field is formed:
\begin{equation}
  B_\tau(\vect{r}) =
  \begin{cases} 1 & \text{if } g(\vect{r}) \geq \tau, \\
                0 & \text{otherwise,} \end{cases}
\end{equation}
where $g(\vect{r})$ is the raw GRAF 1-h precipitation at grid point
$\vect{r}$.  The binary field is then convolved with a Gaussian
kernel of standard deviation $\sigma$:
\begin{equation}
  P_\tau^\sigma(\vect{r}) =
    \frac{\displaystyle\sum_{\vect{r}'} B_\tau(\vect{r}')
      \exp\!\!\left(-\tfrac{|\vect{r}-\vect{r}'|^2}{2\sigma^2}\right)}
         {\displaystyle\sum_{\vect{r}'}
      \exp\!\!\left(-\tfrac{|\vect{r}-\vect{r}'|^2}{2\sigma^2}\right)},
  \label{eq:gauss_prob}
\end{equation}
yielding the Gaussian-smoothed neighborhood probability at each
grid point.  Thresholds of $\tau \in \{0.25, 1.0, 5.0, 10.0\}$\,mm
are evaluated.

The smoothing standard deviation $\sigma$ controls the trade-off
between sharpness and calibration: small $\sigma$ preserves the
spatial detail of the GRAF forecast but yields overconfident
probabilities clustered near 0 or 1; large $\sigma$ improves
reliability at the expense of spatial resolution.  To identify the
optimal $\sigma$, a range of values
$\sigma \in \{3, 5, 7, 10, 15, 20, 25, 30, 40, 50, 60, 75\}$ grid points
(GRAF's native grid spacing is $\approx$4.3\,km)
were evaluated.  For each $\sigma$, MRMS
observations over a held-out verification period were used to
compute reliability diagrams and the Brier Score across all
thresholds.  The value of $\sigma$ that produced approximately the
best-calibrated reliability curves and the lowest aggregate Brier
Score was adopted as the reference standard for subsequent
comparisons with the neural network.

Using the Gaussian-smoothed GRAF field as a reference is preferable
to a raw GRAF exceedance frequency (which would be either 0 or 1
at any single grid point) and avoids the need for ensemble output,
which is not available in this single-deterministic-run context.
The reference standard therefore reflects the information content
of the GRAF forecast itself, and any skill improvement demonstrated
by the neural network represents genuine value added by the
post-processing.

\begin{figure}[H]
\centering
\includegraphics[width=0.75\textwidth]{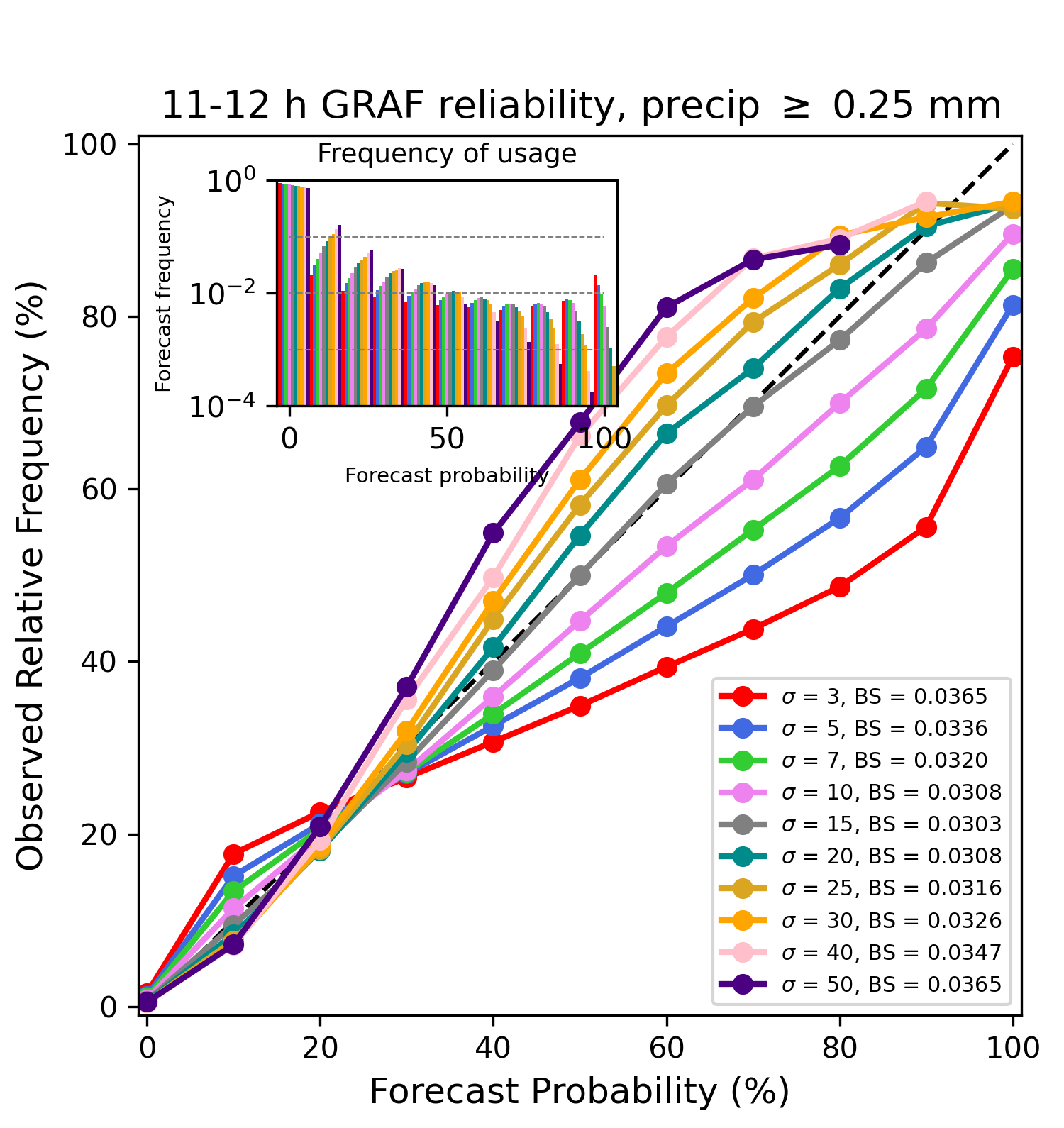}
\caption{Reliability diagrams for Gaussian-smoothed raw GRAF
  neighborhood probabilities of exceeding 0.25\,mm at 12\,h lead
  time, evaluated over January--December 2025.  Each curve
  corresponds to a different Gaussian smoothing standard deviation
  $\sigma$ (grid points; GRAF's native grid spacing is $\approx$4.3\,km); Brier Score (BS) for each $\sigma$ is shown in the
  legend.  The diagonal dashed line indicates perfect reliability.
  }
\label{fig:relia_sigma}
\end{figure}

Figure~\ref{fig:relia_sigma} shows reliability diagrams for the 12-h lead
time and the 0.25-mm threshold using Jan-Dec 2025 and Gaussian-smoothed GRAF forecasts. 
For this lead time and threshold, the selected value $\sigma = 15$ grid points (Table~\ref{tab:sigma})
produces the lowest BS and the most nearly diagonal reliability
curve at this lead time.   Evaluating other thresholds and lead times, approximately
optimal $\sigma$ were chosen, shown in Table~\ref{tab:sigma}.  Gaussian-smoothed probability
forecasts using these thresholds then become the reference standard for comparisons with
the Attention ResUNet.

\begin{table}[ht]
\centering
\caption{Gaussian smoothing standard deviation $\sigma$ adopted for
  the raw-GRAF reference probability forecasts as a function of
  forecast lead time.  $\sigma$ is given in grid points on GRAF's
  native $\approx$4.3-km grid.  Values were selected to minimize the
  aggregate Brier Score across thresholds $\tau \in \{0.25, 1.0,
  5.0, 10.0\}$\,mm over the 2025 verification period.}
\label{tab:sigma}
\begin{tabular}{lc}
\toprule
Lead time (h) & $\sigma$ (grid points) \\
\midrule
1--12    & 15 \\
13--30   & 20 \\
31--48   & 25 \\
\bottomrule
\end{tabular}
\end{table}

\section{Verification metrics}
\label{sec:verification}

Forecast quality is assessed using MRMS (Multi-Radar Multi-Sensor)
radar-derived precipitation as the verification truth.  Verification only uses samples where
the MRMS hourly quality index is $\geq$ 0.5.   This will filter out many points in mountainous
regions where there may be, for example, radar beam blockage.  Standard
verification metrics are described in \citet{wilks2011} and include:
\begin{itemize}
  \item \textbf{Reliability diagrams:} Plotting observed frequency
        against forecast probability for each threshold to assess
        calibration.
  \item \textbf{Brier Skill Score (BSS):} Measuring  the fractional improvement
        over climatological probabilities. 1 is the skill of a perfect forecast, 0 is the
        skill of the reference climatology.
  \item \textbf{Performance diagrams:} Jointly displaying the probability of detection
        and success ratio (1 minus the false alarm ratio) on a single set of axes, with
        shaded background contours of the critical success index and dashed lines of
        constant frequency bias \citep{roebber2009}.
\end{itemize}

\section{Results}
\label{sec:results}

\begin{figure}[H]
\centering
\includegraphics[width=\textwidth]{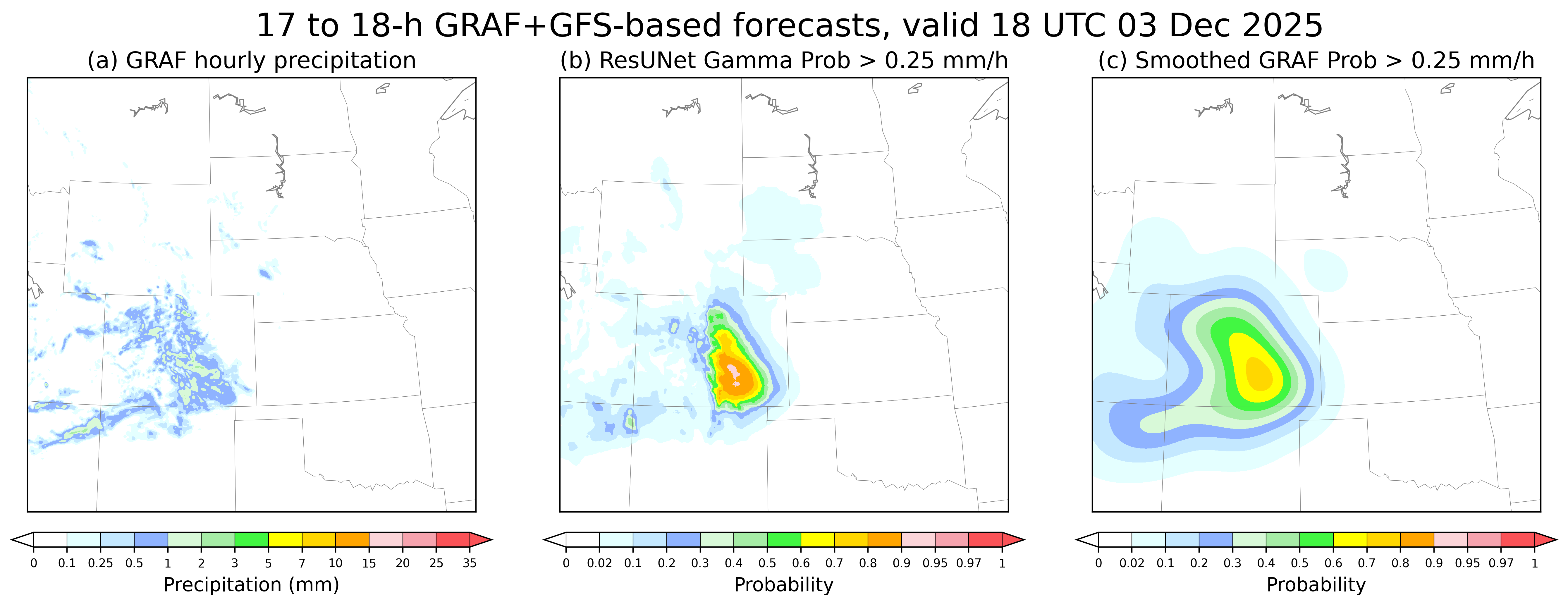}
\caption{Colorado/New Mexico Front-Range upslope snowstorm case.  (a) Raw GRAF 1-h
  accumulated precipitation forecast, (b) Attention ResUNet probability of exceeding
  0.25\,mm, and (c) Gaussian-smoothed GRAF reference probability. }
\label{fig:co_upslope}
\end{figure}

\begin{figure}[H]
\centering
\includegraphics[width=\textwidth]{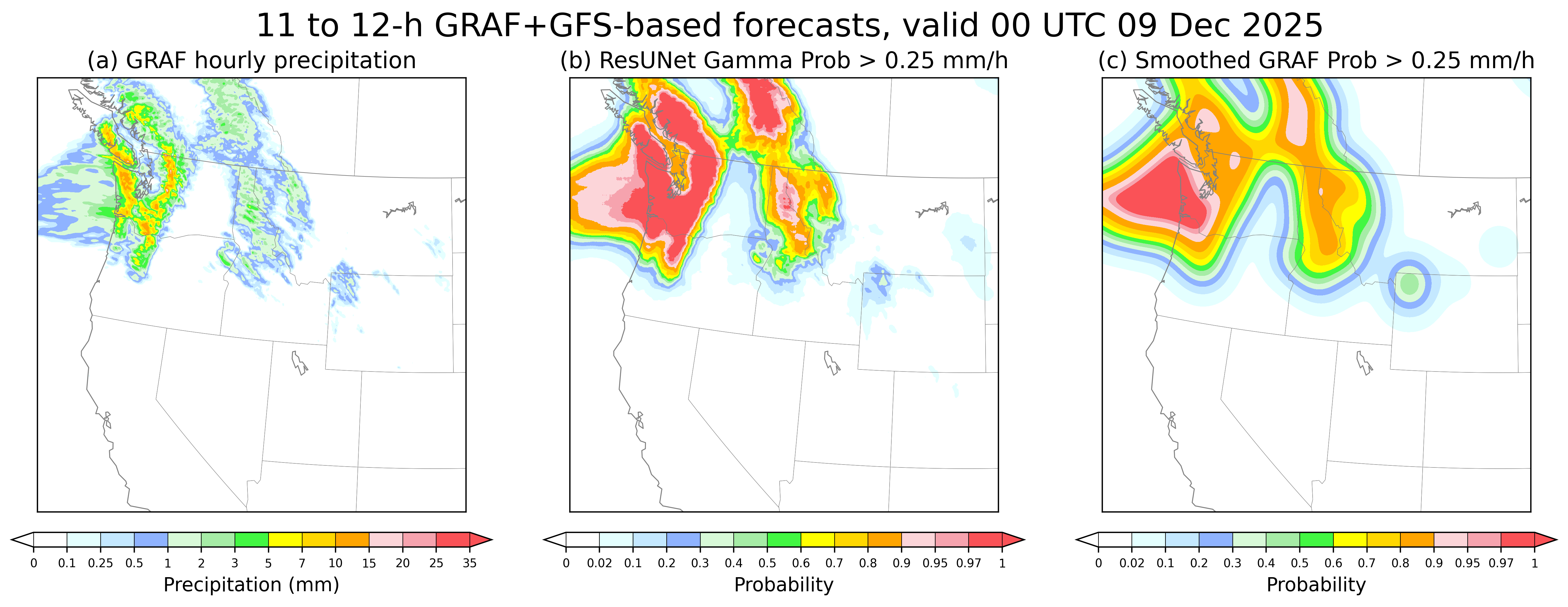}
\caption{Atmospheric river case for Washington State.  (a) Raw GRAF 1-h accumulated
  precipitation forecast, (b) Attention ResUNet probability of exceeding 0.25\,mm, and
  (c) Gaussian-smoothed GRAF reference probability.}
\label{fig:washington_ar}
\end{figure}

\begin{figure}[H]
\centering
\includegraphics[width=\textwidth]{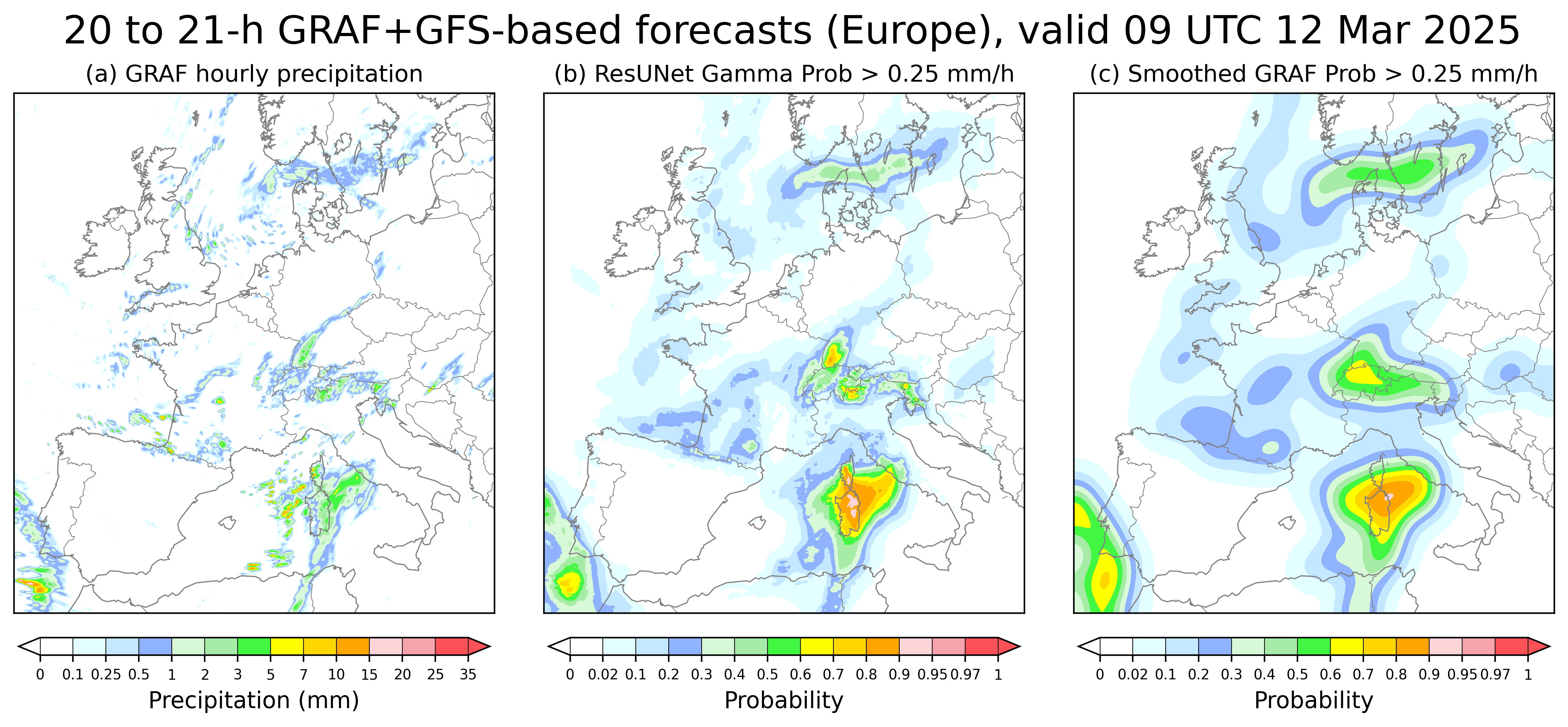}
\caption{Example Attention ResUNet forecast over a portion of Europe.  (a) Raw GRAF
  1-h accumulated precipitation forecast, (b) Attention ResUNet probability of
  exceeding 0.25\,mm, and (c) Gaussian-smoothed GRAF reference probability.}
\label{fig:europe}
\end{figure}

We start by examining spatial forecast maps for realism.   One would expect in these maps that
we would see the probabilities in the Attention ResUNet to be linked to terrain variations and
to incorporate information not just at individual grid points but from surrounding regions.
Figures~\ref{fig:co_upslope}, \ref{fig:washington_ar}, and \ref{fig:europe} provide
representative forecast samples.

Figure~\ref{fig:co_upslope} shows a Colorado/New Mexico Front-Range upslope snowstorm case.  The
Attention ResUNet probabilities in panel (b) show that the largest probabilities are on the
eastern side of the Front Range, whereas the smoothed GRAF based probabilities in panel
(c) include moderate probabilities on the western side of the Front Range.

Figure~\ref{fig:washington_ar} shows an atmospheric river case for Washington State. High probabilities in
the Attention ResUNet are more closely linked to the Olympic Peninsula and the Cascade
Range, whereas the probability maximum for the smoothed GRAF probabilities is more
centered offshore, despite precipitation being forecast to be lighter offshore.   This reflects
the algorithm performing a sequential binary thresholding and smoothing, not accounting
for the amount of precipitation above the threshold.

Could the algorithm be applied in Europe, where GRAF also has a 4-km grid?   While
we do not provide objective verification statistics for Europe in the results presented, 
Fig.~\ref{fig:europe} shows an example of a transfer-learning \citep{pan2010} forecast
in Europe using the US-based training.  Notice the linkage of higher probabilities to the
Italian/Swiss Alps and to the hills in eastern Sardinia, details that are more diffuse with
smoothed GRAF.

\begin{figure}[H]
\centering
\includegraphics[width=\textwidth]{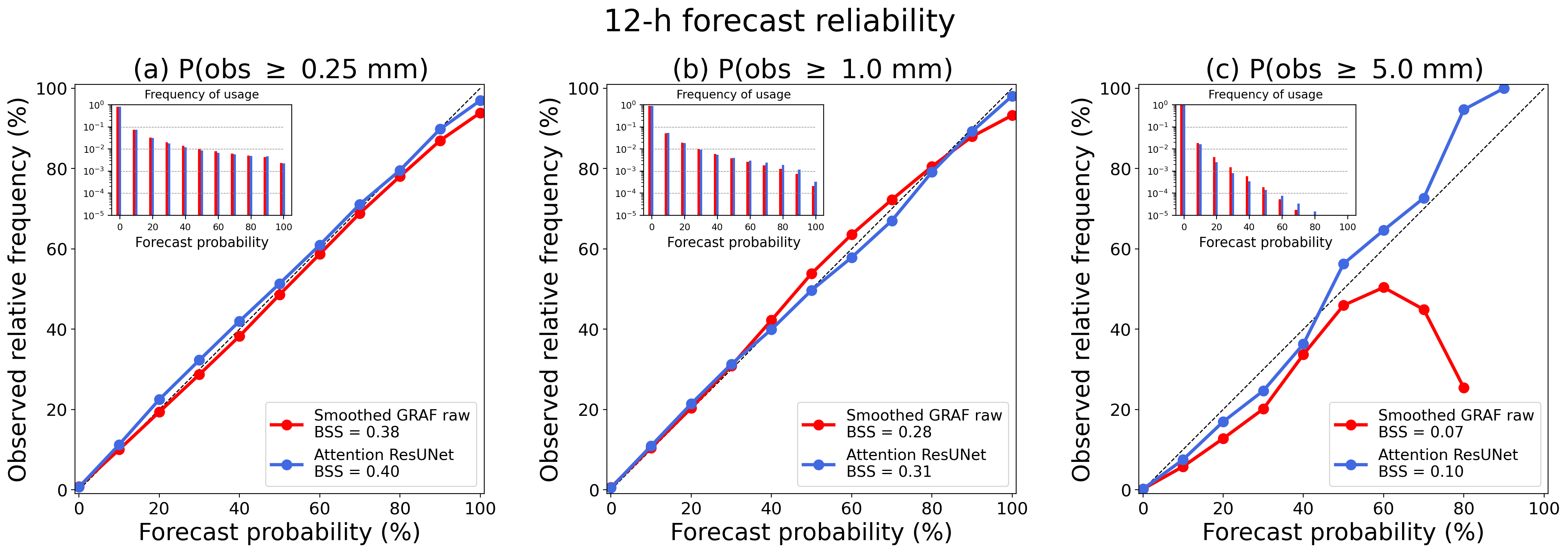}
\caption{Reliability diagrams for the Attention ResUNet (blue) and Gaussian-smoothed GRAF
  reference (red) at 12-h lead time, evaluated over March, June, September, and December 2025.
  Panels show the probability of exceeding (a) 0.25, (b) 1.0, and (c) 5.0\,mm.  Brier Skill
  Scores relative to a Stage IV climatological reference are shown in the legend.  The inset
  histogram shows the frequency of use for each probability bin.  The diagonal dashed line
  indicates perfect reliability.}
\label{fig:relia_12h}
\end{figure}

\begin{figure}[H]
\centering
\includegraphics[width=\textwidth]{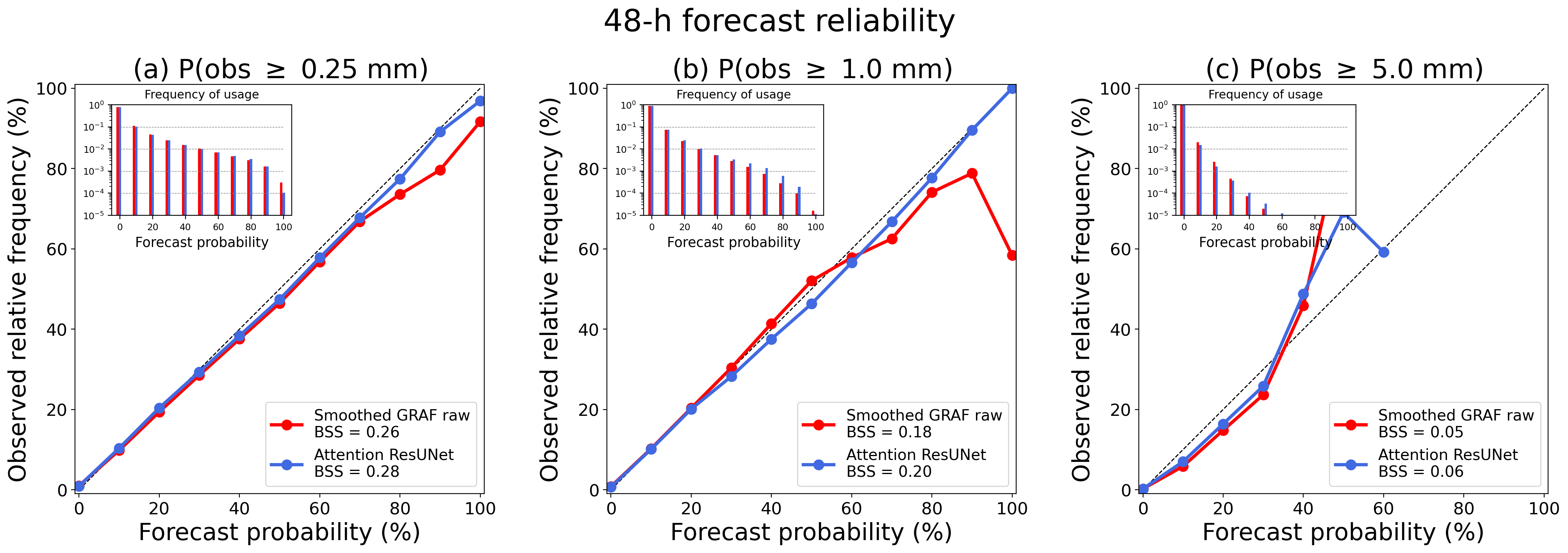}
\caption{As in Fig.~\ref{fig:relia_12h}, but for 48-h lead time.}
\label{fig:relia_48h}
\end{figure}

Figures~\ref{fig:relia_12h} and \ref{fig:relia_48h} provide reliability diagrams for
forecasts of +12\,h and +48\,h, with the panels therein evaluating precipitation exceeding
0.25\,mm (0.01 inches, the US probability of precipitation threshold), 1.0\,mm, and 5.0\,mm.
The reliability diagrams were generated from GRAF deterministic forecasts during the months of March, June,
September, and December 2025, initialized every 6 hours.  Forecasts were evaluated only if
the verifying MRMS analysis had quality that exceeded 0.5.
Brier Skill Scores \citep{wilks2011} are calculated with respect to a climatological reference
standard provided by a climatology generated from the National Centers for Environmental Prediction (NCEP) Stage IV \citep{lin2005} hourly
high-resolution precipitation analyses using 2020--2024 data.  The Stage IV data were
bilinearly interpolated to the GRAF grid before computations of skill.  Reliability diagrams
had largely similar characteristics for other lead times (not shown).  These diagrams show
that generally both methods are capable of providing reliable hourly forecasts.  From the
inset frequency-of-use histograms, for the higher thresholds we see that higher-probability
forecasts are issued more often from the Attention ResUNet.

\begin{figure}[H]
\centering
\includegraphics[width=0.7\textwidth]{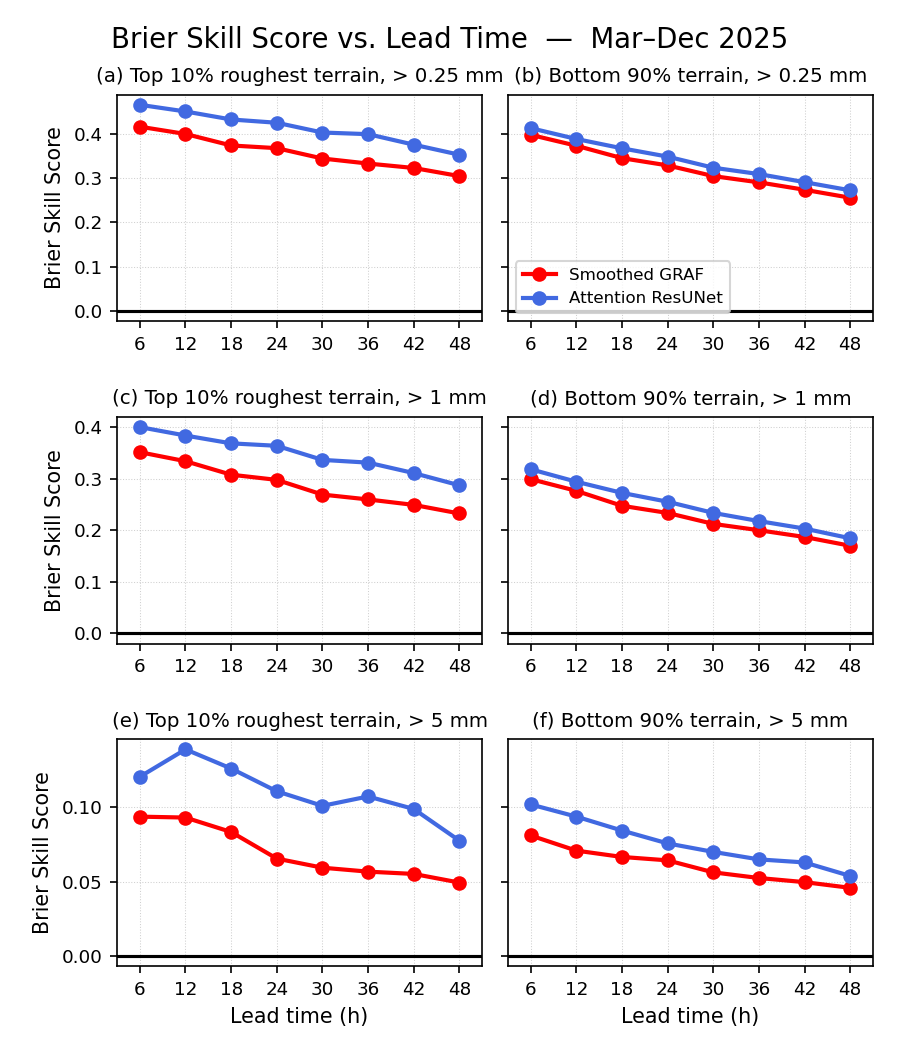}
\caption{Brier Skill Score as a function of lead time for the Attention ResUNet (blue) and
  Gaussian-smoothed GRAF reference (red), stratified by local terrain roughness: (a,c,e)
  the top 10\% roughest terrain and (b,d,f) the remaining bottom 90\% of terrain, as defined
  in the text.  Rows show results for precipitation thresholds of (a,b) 0.25\,mm, (c,d)
  1.0\,mm, and (e,f) 5.0\,mm.}
\label{fig:bss}
\end{figure}

Figure~\ref{fig:bss} shows the Brier Skill scores as a function of lead time, now stratified
by local terrain roughness rather than by CONUS/western-US geography.  Roughness is defined
from the GRAF terrain-height field as a local standard deviation, computed using the
Gaussian-moments identity $\sigma_{\mathrm{local}} = \sqrt{G_\sigma(h^2) - G_\sigma(h)^2}$,
where $h$ is terrain height and $G_\sigma$ denotes Gaussian smoothing with $\sigma = 14$ grid
points ($\approx$60\,km on GRAF's $\sim$4.3-km grid).  The ``top 10\%'' terrain category
comprises the grid points whose local standard deviation exceeds the 90th percentile of this
quantity computed over all finite terrain points in the GRAF domain; the remaining points form
the ``bottom 90\%'' category.  This is a substantially more direct measure of complex terrain
than the West-of-105$^{\circ}$W criterion used earlier, and it reveals that the Attention
ResUNet's advantage over the Gaussian-smoothed GRAF reference is concentrated in the roughest
terrain: at every lead time and threshold, the BSS gap between the two forecasts is larger in
the top-10\% category than in the bottom-90\% category.  This gap is most pronounced for the
5\,mm exceedance threshold (Figs.~\ref{fig:bss}e,f), where the Attention ResUNet retains a BSS
of roughly 0.10--0.20 over rough terrain through 48\,h, 25--55\% higher than the reference
forecast's skill there depending on lead time, while over smoother terrain the two forecasts
are closer together and both decay to a BSS below 0.10 by 48\,h.  The improvement over the reference forecast is
smaller, though still consistent, for the 0.25 and 1.0\,mm thresholds.

\begin{figure}[H]
\centering
\includegraphics[width=\textwidth]{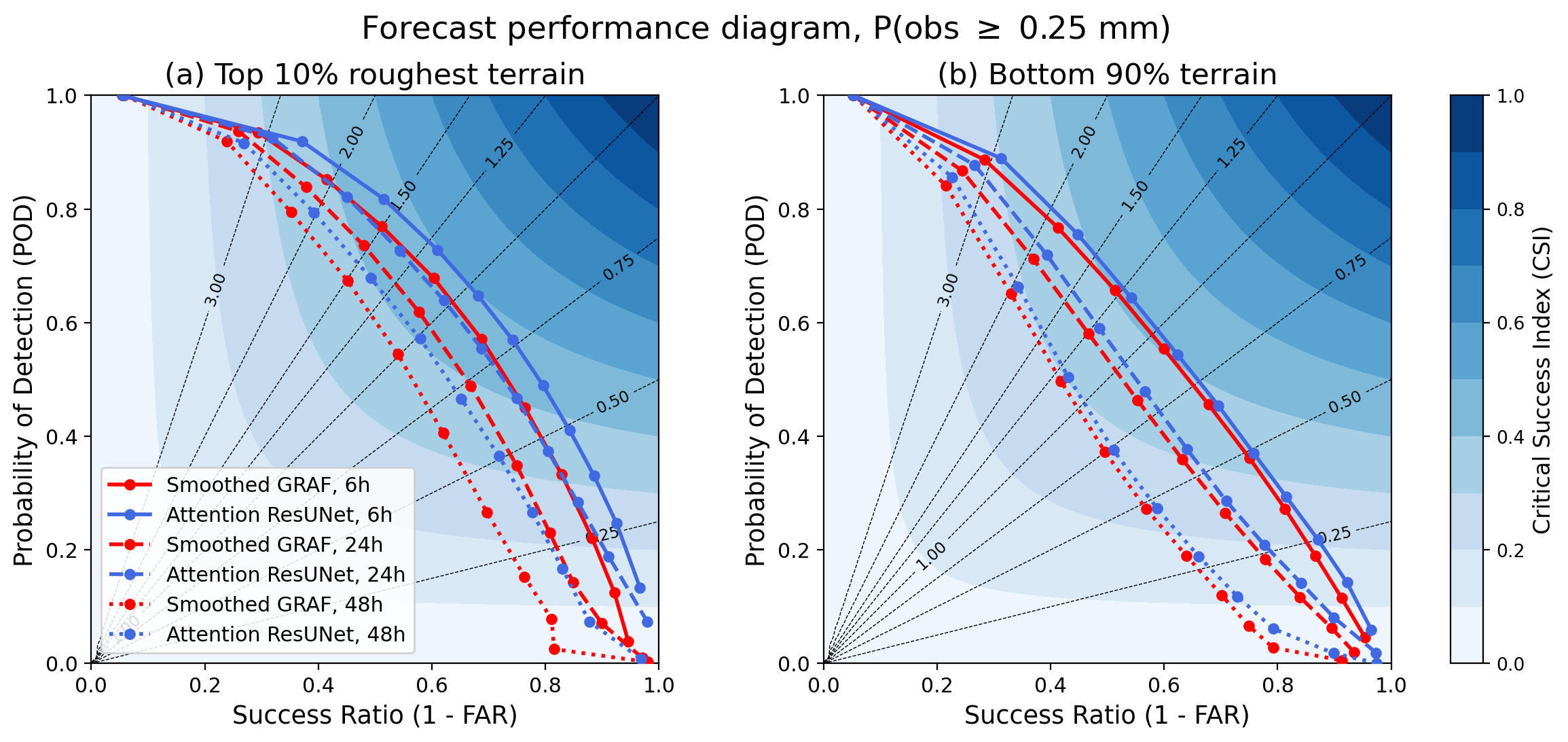}
\caption{Performance diagrams (probability of detection vs.\ success ratio, with shaded
  critical success index and dashed frequency-bias contours) for the 0.25\,mm exceedance
  threshold, for (a) the top 10\% roughest terrain and (b) the bottom 90\% of terrain.  Curves
  show the Attention ResUNet (blue) and Gaussian-smoothed GRAF reference (red) at 6-h (solid),
  24-h (dashed), and 48-h (dotted) lead times, with each curve built by cumulating
  contingency-table counts across forecast-probability bins from 0 to 100\% in 10\% increments.}
\label{fig:perfdiag_p25}
\end{figure}

\begin{figure}[H]
\centering
\includegraphics[width=\textwidth]{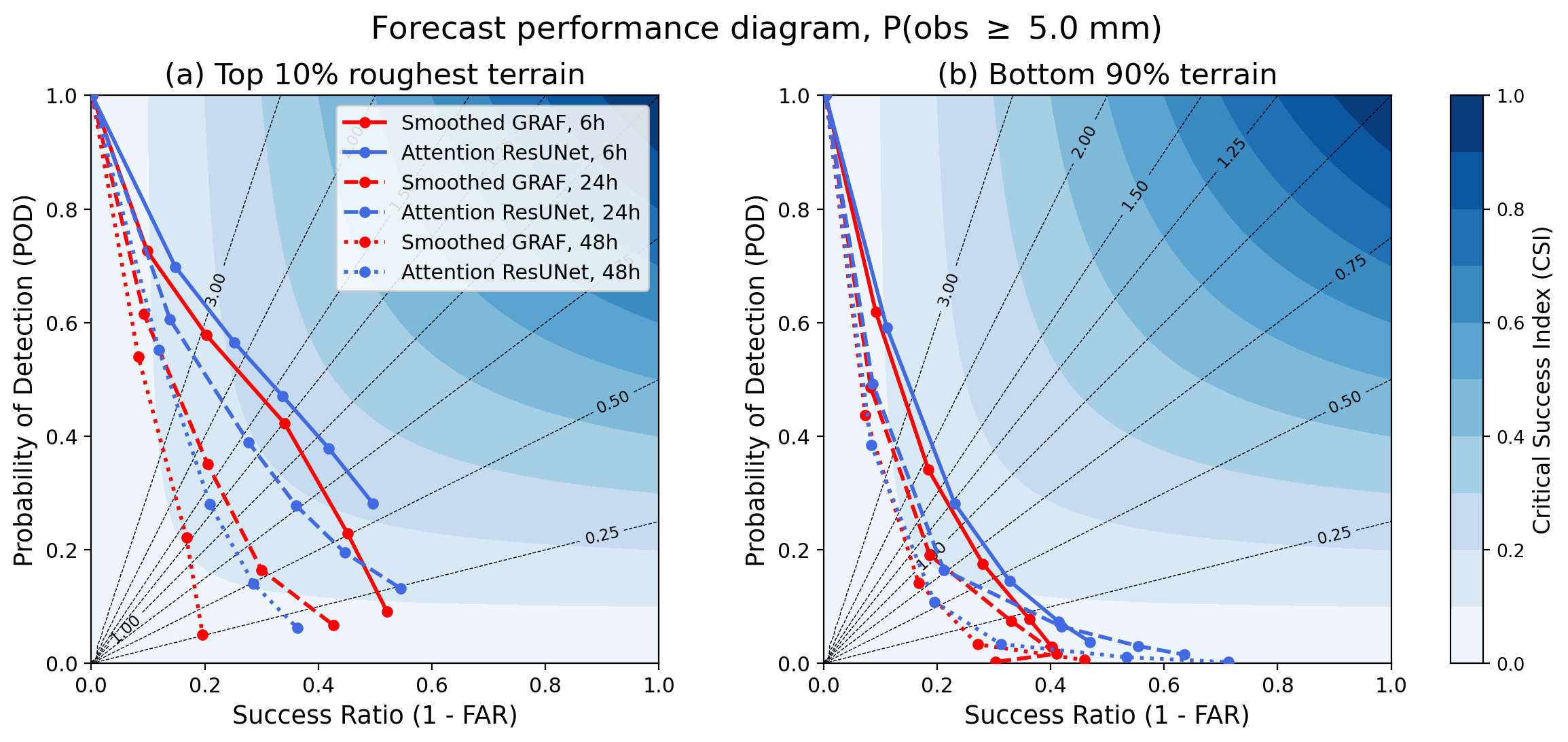}
\caption{As in Fig.~\ref{fig:perfdiag_p25}, but for the 5\,mm exceedance threshold.
  Forecast-probability bins above 50\% are omitted because they are used too rarely at this
  threshold, particularly by the reference forecast over the top-10\% roughest terrain, for
  the resulting POD/SR estimates to be statistically meaningful.}
\label{fig:perfdiag_5mm}
\end{figure}

Performance diagrams \citep{roebber2009} provide a complementary view of forecast quality by
jointly displaying the probability of detection (POD) and success ratio (SR, equal to 1 minus
the false alarm ratio), with shaded background contours of the critical success index (CSI)
and dashed lines of constant frequency bias.  For each of the top-10\% and bottom-90\% terrain
categories defined above, curves are constructed by treating each of the 11 saved
forecast-probability bins (0, 10, 20, \ldots, 100\%) as a probability threshold above which
the forecast is counted as ``yes,'' and cumulating hit, false-alarm, and miss counts from that
bin through the highest bin.  Figures~\ref{fig:perfdiag_p25} and \ref{fig:perfdiag_5mm} show
these diagrams for the 0.25 and 5\,mm thresholds, respectively, with curves for the Attention
ResUNet and Gaussian-smoothed GRAF reference at 6-, 24-, and 48-h lead times plotted together
(distinguished by line style) on the same axes.  For the 0.25\,mm threshold, both forecasts
trace out smooth, well-behaved curves, with the Attention ResUNet consistently shifted toward
higher POD and higher SR (i.e., toward the upper right, higher-CSI corner of the diagram) than
the reference forecast at every lead time, and with the top-10\% terrain category again
showing the larger separation between the two forecasts.  For the 5\,mm threshold, the highest
forecast-probability bins (above 50\%) are used so rarely, particularly by the reference
forecast over the top-10\% roughest terrain, that the resulting points are dominated by
sampling noise and produce a non-monotonic curve; those points (bins above 50\% forecast
probability) are therefore omitted from Fig.~\ref{fig:perfdiag_5mm}.  Even with this
truncation, the same pattern seen in the BSS results is evident: the Attention ResUNet
maintains a higher POD than the reference forecast for a given SR at all three lead times,
and the advantage is again largest over the top-10\% roughest terrain.

\section{Summary}
\label{sec:summary}

We have described a probabilistic precipitation forecasting algorithm for the statistical
postprocessing of deterministic forecast model output, in this case applied to 4-km deterministic
forecast guidance produced by TWCo's GRAF model, based on the MPAS system. This method
uses an Attention Residual U-Net to produce at each point a probability distribution
from a zero-inflated mixture of two Gamma distributions.  Key design choices include:
\begin{itemize}
  \item A \emph{two-component Gamma mixture} that accommodates both
        stratiform and convective precipitation modes within a
        single parametric framework, with a hard shape-ordering
        constraint to prevent label switching between components.
  \item \emph{Attention gates} that allow the decoder to focus on
        the most informative encoder features at each spatial
        location, suppressing irrelevant dry-region activations
        and providing orographically selective skip connections.
  \item A \emph{power transformation} of the input precipitation
        field for variance stabilization.
  \item \emph{EM-based climatological initialization} of the output
        layer biases, fitting a two-component Gamma mixture to wet
        training pixels for stable, physically grounded convergence.
  \item \emph{Single-pass full-domain inference}: the fully
        convolutional network processes the entire CONUS domain in
        a single forward pass after edge-replication padding,
        eliminating patch-boundary artifacts.
  \item A \emph{Gaussian-smoothed GRAF reference standard} produced
        by convolving the raw deterministic forecast's binary
        exceedance field with the optimal Gaussian kernel, providing
        a calibrated single-model baseline for skill assessment.
\end{itemize}

The NLL loss function is a proper scoring rule, ensuring that the
model is incentivized to produce well-calibrated and sharp
probabilistic forecasts.

Several future improvements of this basic system and further validation are possible
but were not performed here.   The statistical characteristics of precipitation forecasts over
Europe were not examined, in part because there is no one convenient data set that we
are aware of with the characteristics of the US MRMS data, synthesizing model, radar, and
surface observations, with extensive quality control.

Another deficiency is that because of the limited GRAF data saved (hourly forecast precipitation
four times daily, but no other forecast fields, nor data from the every-hour production of GRAF), we
were missing other helpful predictive features such as column relative humidity directly from GRAF, or
the use of lagged ensemble forecasts to improve the potential predictive skill, understanding
what features were consistent from run to run and which differed.  These deficiencies,
time permitting, may be addressed in future work, or the method may be applied to other models
with richer training data sets.

Finally, it's likely the training procedure could be improved,
for example with training methods like FiLM \citep{perez2017} to still
extract lead-time dependent skill with training data pooled over multiple lead times.

\section*{Acknowledgments}

Trevor Alcott is thanked for his comprehensive internal review of this manuscript.



\end{document}